\documentclass{emulateapj}
\usepackage{pslatex}
\def\idm#1{{\mbox{\scriptsize #1}}}

\newcommand\Ym{\langle Y\rangle}
\newcommand\sm{\langle \sigma\rangle}
\newcommand\Chi{{(\chi^2_\nu)^{1/2}}}

\def\url#1{\texttt{#1}}

\def\idm#1{{\mbox{\scriptsize #1}}}
\newcommand\muarae{{$\mu$~Ara}}
\newcommand\hd{{HD160691}}

\begin{document}

%
%------------------------------------------------------------------------------
%
%
\title{On the extrasolar multi-planet system around HD~160691}
%
%------------------------------------------------------------------------------
%

\author{Krzysztof Go\'zdziewski\altaffilmark{1},
        Andrzej J. Maciejewski\altaffilmark{2}, and
        Cezary Migaszewski\altaffilmark{1}
       }
\altaffiltext{1}{Toru\'n Centre for Astronomy,
  N.~Copernicus University,
  Gagarina 11, 87-100 Toru\'n, Poland; k.gozdziewski@astri.uni.torun.pl}
\altaffiltext{2}{Institute of Astronomy, University of Zielona G\'ora,
  Podg\'orna 50, 65-246 Zielona G\'ora, Poland; maciejka@astro.ia.uz.zgora.pl}
%
%------------------------------------------------------------------------------
%
\begin{abstract}
We re-analyze the precision radial velocity (RV) observations of  \hd{}
(\muarae{})  by the Anglo-Australian Planet Search Team. The star is supposed to
host two Jovian companions (\hd{}b, \hd{}c) in long-period orbits ($\sim
630$~days and  $\sim 2500$~days, respectively) and a  hot-Neptune (\hd{}d) in
$\sim 9$~days orbit. We perform a global search for the best fits in the orbital
parameters space with a hybrid code employing the genetic algorithm and simplex
method. The stability of  Keplerian fits is verified with the $N$-body model of
the RV signal that takes into  account the dynamical constraints (so called GAMP
method). Our analysis reveals a signature of the fourth, yet  unconfirmed,
Jupiter-like planet \hd{}e in $\sim 307$~days orbit. Overall, the global
architecture of four-planet  configuration recalls the Solar system. All
companions of \muarae{} move in quasi-circular orbits.  The orbits of two inner
Jovian planets are close to the 2:1~mean motion resonance. The alternative
three-planet system involves two Jovian planets in eccentric orbits ($e\sim
0.3$), close to the 4:1~MMR, but it yields a significantly worse fit to the
data. We also verify a hypothesis of the 1:1~MMR in the subsystem of two inner
Jovian planets in the four-planet model.
\end{abstract}
%
%------------------------------------------------------------------------------
%
%
%------------------------------------------------------------------------------
%
\section{Introduction}
%
%______________________________________________________________________________
%
The star \hd{} (\muarae{}) is a Sun-like main-sequence dwarf monitored by the
long-term, precision radial velocity (RV) surveys.  It has been observed over
more than 7~years by the Anglo-Australian Telescope (AAT)  Planet Search Team
and by the Geneva Planet Search Team (CORALIE and HARPS spectrometers).  The
work of the AAT team lead to the discovery of a Jupiter like companion \hd{}b in
about 630~days orbit \citep{Butler2001}. One year later, \cite{Jones2002a}
confirmed the Jovian planet and discovered a linear trend in the RV data
revealing  a signature of the second, more distant body. In the next paper, 
\cite{McCarthy2004} published a new  orbital solution with the orbital period of
the long-period planet \hd{}c about 3000~days and the eccentricity $e_{\idm{c}}
\sim 0.57$. The same year, \cite{Santos2004}, using observations done with the 
ultra precise HARPS spectrometer,  announced $\sim 14$ Earth-mass planet \hd{}d
in $\sim 9$~days orbit.  That discovery is a breakthrough in the field as the
long-term precision of spectrometers approaches $1$~m/s. Actually, the
instrumental errors are much smaller than the RV  variability (stellar jitter)
induced by the Sun-like stars themselves. Recently, \cite{Butler2006} published
a new updated set of 108 observations of \muarae{}, spanning 2551~days (about
7.5~yr). Thanks to the  updates in the UCLES instrument installed at the AAT and
the software pipeline \citep{Butler2006}, the long-term precision of the
measurements is amazing. It also reaches  $1$~m/s at the end of the
observational window and is kept at the mean level of $\sim 2.8$~m/s over its
whole length. 

This paper is devoted to the analysis of the RV observations of \muarae{} using
the so called GAMP approach (an acronym of genetic algorithm with MEGNO penalty)
that makes explicit the use of Newtonian, self-consistent $N$-body model of the
stars' reflex motion, dynamical properties of the planetary system and the
Copernican Principle \citep{Gozdziewski2003e,Gozdziewski2005a}. This algorithm
makes it possible to  derive meaningful bounds on the elements of the outermost
planet in spite of the fact that the data cover only a part of the  longest
orbital period. Without the dynamical constrains, the kinematical as well as
pure Newtonian best fits to the \muarae{} RV tend to show large eccentricity of
the outermost companion and then the system becomes catastrophically  unstable. 

We verify the results of the previous papers based on a much smaller number of
relatively less accurate RV observations.  The results of the analysis of the
new RV data set greatly improved and extended over time by \cite{Butler2006}
lead us to a conclusion that the \muarae{} may host {\em four} planets in
quasi-circular orbits, including a new, Jupiter-like object in $\sim 307$~days
orbit (\hd{}e)\footnote{In this paper we use a widely adopted (but unofficial)
convention for naming the extrasolar planets with lower-case Roman  letters
starting from ``b'', in the order of their announcement.}. The new best fit
solution describes the orbital architecture of this extrasolar system as very
different from the previous ones.

Shortly after submitting the manuscript, we learned about an independent work by
\cite{Pepe2006} who announced  a very similar orbital solution and  also the
fourth planet in the \muarae{} system. These authors study a  different set of
the RV measurements but the conclusions of the two papers are in an excellent
agreement. Still, in this paper we focus our attention on the analysis of the
AAT data along  the lines of the originally submitted manuscript. However, the
last Section~5 is  devoted to a preliminary study of the full available data
set, including the new  measurements published by \cite{Pepe2006}. In
particular, we verify a hypothesis  of the 1:1~MMR in the subsystem of two
Jovian planets. %
%------------------------------------------------------------------------------
%
\section{Fitting multi-planet configurations to RV data}
%
%______________________________________________________________________________
%
According to the previous papers devoted to \muarae{},  the time range of the 
updated data set published by \cite{Butler2006} should be already close to the
orbital period of the outermost companion. In that case, we can try to recover a
good approximation of the system parameters by modeling the RV signal with the
Keplerian orbits. Although for mutually interacting systems the kinematic model
often leads to unstable orbital configurations, thanks to the numerical
simplicity it can be  very helpful to rapidly determine  putative solutions and
interesting  ranges of orbital parameters. 

The contribution of every planet to the reflex motion of the parent star at time
$t$ is the following  \cite{Smart1949}: 
\begin{equation} 
V_{\idm{r}}(t) = K [ \cos (\omega+\nu(t)) + e \cos \omega] + V_0, \label{eq:eq1}  
\end{equation} 
where $K$ is the semi-amplitude, $\omega$  the argument of pericenter, $\nu(t)$ 
the true anomaly involving implicit dependence on the orbital period  $P$ and
the time of periastron passage $T_{\idm{p}}$, $e$  the eccentricity, and $V_0$
is the velocity offset. Some argue that it is best to interpret the derived fit
parameters $(K,P,e,\omega,T_{\idm{p}})$ in terms of  Keplerian elements and
minimal masses related to Jacobi coordinates \citep{Lee2003,Gozdziewski2003e}.
We follow their reasoning when calculating the orbital elements from the primary
fit parameters.

The extensive exploration of the multi-parameter $\Chi$ space is  efficient
enough if one applies a kind of a hybrid optimization  \citep{Gozdziewski2004,
Gozdziewski2006b}. The idea of this algorithm relies on two steps, global and
local ones. In the first step, we search for potentially good (but not very
accurate) solutions in a global manner. During the second step these solutions
become initial conditions to a precise and fast local algorithm. The single
program run starts the genetic algorithm (GAs). In particular, we apply the
PIKAIA code by \cite{Charbonneau1995}. The GAs have important advantages over
more popular gradient-type methods \citep{Press1992}. The power of GAs lies in
their basically global nature,  the requirement of knowing only the $\Chi$
function, and the ease of constrained optimization; GAs permit defining
parameter bounds according to specific requirements, or adding a penalty term to
$\Chi$ \citep{Gozdziewski2006a}.  However, the best fits found with GAs are (in
principle) not very accurate in terms of $\Chi$ or rms, so we refine them with
another non-gradient algorithm,  the simplex method of Melder and
Nead~\citep{Press1992}.  Usually,  we run such hybrid procedure thousands of
times, and then we analyse the ensemble of gathered fits. That helps us to
detect local minima of $\Chi$ that are sometimes very distant in the parameter
space and to get reliable approximation of the global topology of $\Chi$. We
tested the code extensively \citep{Gozdziewski2006b}, and we found many examples
confirming its robustness and reliability.  Remarkably,  the code works with 
minimal requirements for user-supplied information:  basically, one should only
define the model  function [so called {\em fitness function}, usually equal to
$1/\Chi$] --- conveniently, it is the same for the GAs and simplex, and to
determine (even very roughly) parameter bounds for the assumed number of
planets. We underline that the code is FFT-free, and by its construction,  it
works without any {\em a-priori} determination of the orbital periods.

In some cases (like strongly resonant or interacting  systems, noisy data, a
small number of measurements) the kinematic fits may lead to unrealistic, 
rapidly disrupting configurations.  Then a more elaborate $N$-body Newtonian
model of the RV should be applied \citep{Rivera2001,Laughlin2001}.  Yet the
hybrid optimization can be still used as a general approach of exploring the
parameter space (only the model function is changed). Actually, even more
general modeling of the RV data relies on the elimination of the  unstable
(strongly chaotic) solutions {\em during} the fit process. That  self-consistent
approach follows the Copernican Principle and the  complex structure of the
phase space predicted by the Kolmogorov-Arnold-Moser theorem
\citep{Arnold1978}.  The idea of the dynamical fits  with stability constraints
relies on modifying the $\Chi$ function by a penalty term employing an efficient
fast indicator MEGNO \citep[]{Cincotta2000}. We describe that method (GAMP) in
detail in our past works \citep{Gozdziewski2005a,Gozdziewski2006a}.

To obtain reliable estimates of the fitted parameter errors, the internal
measurement errors should be rescaled \citep{Butler2004} according to
$
 \sigma^2 = \sigma_{\idm{m}}^2 + \sigma_{\idm{j}}^2,
$ 
where  $\sigma_{\idm{m}}$ and $\sigma_{\idm{j}}$  is the internal error and
adopted dispersion of stellar jitter, respectively, and $\sigma$ is the joint
uncertainty. Typically, we choose $\sigma_{\idm{j}}$ following the estimates for
Sun-like dwarfs by \cite{Wright2005}, or we use the value adopted by the
discovery teams. The jitter estimate for \muarae{} by \cite{Butler2006} is
3.5~m/s and we use this value in our calculations.  The analysis of the short
time-scale RV variability of the parent star may be found in \cite{Bouchy2005}.
The physical properties of the parent star are discussed in \cite{McCarthy2004}
and \cite{Santos2004}.

%
%------------------------------------------------------------------------------
%
\section{Three-planet model of the RV}
%
%------------------------------------------------------------------------------
%
In \citep{Gozdziewski2003e,Gozdziewski2005a} we carried out an extensive
analysis of the RV measurements published by \cite{Jones2002a} and
\cite{McCarthy2004} assuming that there exist two Jupiter companions of
\muarae{}.  By employing the dynamical $N$-body model and GAMP, we obtained
meaningful limits on the barely constrained orbital parameters of the putative
outermost planet. According to our results,  $a_{\idm{c}}$  should be roughly
greater that $4$~AU and $e_{\idm{c}}<0.4$ in the range of the smallest
permissible  semi-major axes. The new precision RV data published by
\cite{Butler2006} give us an excellent opportunity to verify these conclusions. 

Figure \ref{fig:fig1} shows the results of the hybrid search for the putative
three-planet configurations, in terms of multi-planet Keplerian model, assuming
that the innermost planet has $P_{\idm{d}}\in [7,12]$~days and $e_{\idm{d}}\in
[0,0.3]$, and that two Jupiter-like planets are in orbits with $P_{\idm{b}}\in
[100,1200]$~days,  $P_{\idm{c}}\in [1200,8500]$~days, and $e_{\idm{b,c}} \in
[0.0,0.8]$, respectively.  These safe assumptions are consistent with the
results of a few  papers devoted to \muarae{}. In particular, we rely on the
careful analysis by \cite{Santos2004} and \cite{Bouchy2005} that revealed the
hot-Neptune \hd{}d. We might expect that its signal $K_{\idm{d}} \sim 4$~m/s is
comparable to  the error level of the AAT data, nevertheless we decide to add
this planet to the model, not to avoid the {\em a-priori} information on the
system architecture.  It appears that the hot-Neptune signal improves the rms by
$\sim 0.5$~m/s,  so it could be important to obtain a precise solution for the
whole system.

The best fits obtained in the search are illustrated by  projections onto the
planes of particular parameters of the RV model, Eq.~1. Here, we choose the
$(P,K)$- and $(P,e)$-planes. Marking the elements within the formal
$1\sigma,2\sigma,3\sigma$ confidence intervals of the best-fit solution  (signed
by two crossing lines), we have a convenient way of visualizing the  shape of
the local minima of $\Chi$ and obtaining realistic and reasonable estimates of
the parameter errors \citep{Bevington2003}.  Figure~1 illustrates the
parameters  of $\sim 1000$ different fits within the $3\sigma$ confidence
interval of  the best Fit~I (its parameters are given in Table~1). Remarkably,
the orbital period $P_d \sim 9.637$~days and the semi-amplitude $K_d \sim
3$~m/s  are very close to the independent estimates by \cite{Santos2004}, on the
basis of HARPS measurements.  Thus in spite of the RV contribution of the
hot-Neptune planet~d  (of the inferred $\sim 11$~Earth-masses) being on the
noise level $\sim 4$~m/s, the planet is already ``visible'' also in the AAT 
measurements. The two-planet Keplerian model of the RV 
yields\footnote{The parameters $(K,P,e,\omega,T-T_0)$ in Eq.~1 are
(37.52~m/s, 630.31~days,      0.269,    259.63~deg,  13401.53~days), and
(18.10~m/s,   2499.16~days,      0.466,    184.04~deg,  11032.98~days),
$V_0=-0.03$~m/s, $T_0=$JD2,440,000.}
$\Chi \sim1.11$ and an rms
$\sim4.45$~m/s, so the signal of \hd{}d 
improves the fit by
$\sim  0.5$~m/s. Yet in the next section, we are trying to find  much better arguments
supporting this claim.

In turn, the elements of the Jupiter-like companions seem constrained very
well.  For a reference, the synthetic curve and the data points are illustrated
in Fig.~\ref{fig:fig2}. An rms of Fit~I is $\sim 4$~m/s and its $\Chi\sim 1$.
The best-fit three-planet solution found here is very  similar to the one quoted
by  \cite{Butler2006}, see their Table~3.  The orbital period ratio close to 4:1
suggests a proximity of the Jovian planets to the 4:1~mean motion resonance
(MMR). Unfortunately, the system is again catastrophically unstable due to a
large  eccentricity of the outer planet ($e_{\idm{c}}\sim {0.47})$ and  a
proximity of both orbits to the collision zone, i.e., the area close to the
planetary collision line determined by
$a_{\idm{b}}(1+e_{\idm{b}})=a_{\idm{c}}(1-e_{\idm{c}})$.

However, it is well known that orbits involved in low-order MMRs may be stable
even if  they are crossing each other \citep{Ji2003,Beauge2003,Psychoyos2005}. 
In \cite{Gozdziewski2006a} we  re-analyse the RV of HD~108874 \citep{Vogt2005} 
that appears  to host two Jovian planets very close to the same type of the
4:1~MMR. The dynamical map of this system reveals that the eccentricity of the
outer planet could be as large as 0.7 in the stable resonance island, although
already for $e_{\idm{c}}\sim 0.4$ the osculating orbits would cross. In the same
way, the exact 4:1~MMR could explain large $e_{\idm{c}} \sim 0.5$ in the
three-planet \muarae{} system. To  examine more carefully the stability of the
best-fit configuration, we computed dynamical maps (Fig.~\ref{fig:fig3}) in the
$(a_{\idm{c}},e_{\idm{c}})$-plane,  in terms of the Spectral Number ($SN$)
\citep{Michtchenko2001} and the $\max e$ indicator (the maximal eccentricity
attained during the integration time-span). In particular, every point in these
maps represents  an initial condition that was  integrated over $\sim 10^5$~yr 
($\sim 10^4 P_{\idm{c}}$).  The dynamical maps reveal a few dominant low-order
MMRs: like 4b:1c, 9b:2c and 5b:1c in the neighborhood of Fit~I, marked by a
crossed circle ($n_{\idm{b}}\mbox{b}:n_{\idm{c}}\mbox{c}$ means the
$n_{\idm{b}}:n_{\idm{c}}$~MMR  of planets ``b'' and ``c'').  Clearly, the
best-fit Keplerian configuration lies in a strongly chaotic
zone\footnote{
In \citep{Gozdziewski2005a} we already derived a very similar solution  to the
ones quoted by \cite{Butler2006} and found in this work, with $a_{\idm{c}}\sim
3.8$~AU and large $e_{\idm{c}}\sim 0.6$, on the basis of RV data from
\cite{McCarthy2004} extended by  measurements published graphically in
\cite{Santos2004}. However, we could not find any stable solution in its close
neighborhood.
}.

A simple change of the  Keplerian best-fit $e_{\idm{c}}$ to $\sim 0.25$,
providing a stable system, leads to a significant increase of $\Chi$, so we
tried to find an optimal stable solution with GAMP. Due to a large CPU
requirement caused by the short-period orbit of the innermost low-mass  \hd{}d
that  planet has been skipped in this test.  We searched for the two-planet
solutions only. In the penalty function \citep{Gozdziewski2003e} we integrated
the MEGNO over $\sim 10^3 P_{\idm{c}}$. It is  a relatively short time  but it
enables us to  rule out strongly chaotic (and  rapidly disrupting) systems. The
results of that search are illustrated in Fig.~\ref{fig:fig4} were  projections
of the best-fit parameters onto the dynamical maps in the
$(a_{\idm{c}},e_{\idm{c}})$-plane are shown. Only the solutions within the
$1\sigma$  confidence interval of the best fit (marked by the largest circle,
see also caption to Fig.~\ref{fig:fig4} for its osculating elements) are shown. 
Let us note that this best-fit solution has been refined by GAMP integrations
over $\sim 25,000 P_{\idm{c}}$ and the solution appears to be rigorously stable.
An rms of these two-planet fits is $\sim 4.7$~m/s and their  $\Chi \sim 1.17$.
The scatter of the best-fit parameters is small but we cannot decide whether the
system is locked in the  exact 4:1~MMR. More likely it evolves close to its
separatrix (outside the resonance island). Note that the fits with the largest
$e_{\idm{c}}$ (to the left of the resonance island) are in fact mildly chaotic.
The presence of the Neptune-like companion does not lead to any qualitative
changes in the dynamical character of the best fit  configurations although it
yields a lower rms $\sim 4.4$~m/s (for {\em stable} configurations).

The results of modeling the RV data by the three-planet configurations seem   in
an overall agreement with the conclusions of our previous work 
\citep{Gozdziewski2005a}.  However, we found that the data published in
\cite{McCarthy2004} rather exclude the possibility of a stable 4:1~MMR between
the Jovian planets. The acceptable (stable) solutions should have $a_{\idm{c}}$
roughly not smaller than $4$~AU. The corresponding orbital period is
significantly longer, by $\sim 500$~days, from the current apparently very
precise estimate of $P_{\idm{c}} \sim 2500$~days found on the basis of the new
data set. Still, although the observational window already covers about one
outermost period $P_{\idm{c}}$  and the data strongly constrain $\Chi$, both the
kinematic as well as the Newtonian model of the RV yield catastrophically
unstable orbital configurations. 

That lead us to look for an explanation of the strange inconsistency. The most
natural one could follow  from the existence of an additional planet that has
been hidden up till now due to the small number of measurements and their
significant errors ($\sim 4$~m/s, as quoted in the older  papers by the AAT
Team).  The problem reminds us of the study of the HD~37124 data
\citep{Vogt2005,Gozdziewski2006a}. For this star, the two-planet fits are
strongly unstable due to the extreme  eccentricity of the outer  companion,
$\sim 0.7$. Recently, \cite{Vogt2005} has shown that the assumption of
three-planets  makes it possible to improve the fits and, simultaneously, the
best-fit orbits become close to circular ones. The system can also be easily
stabilized in such a regime \citep{Gozdziewski2006a} without any degradation
$\Chi$ and the rms.  Further, we follow the results of \cite{Bouchy2005} who
detected the short-period planet~d around \muarae{}  with ultra-precise HARPS
observations.   In particular, we are attracted by the analysis of the
short-time scatter of RV during several subsequent nights (see their Table~1 and
Fig.~1,2). That statistics  measures the RV variability imposed by the star
activity itself. The standard deviation of these variations over one night is
between 1.5--2.5~m/s. It could mean that $\sigma_{\idm{j}}$ is in fact less than
3.5~m/s that we adopted here. Instead, assuming $\sigma_{\idm{j}} \sim 2$~m/s we
got $\Chi \sim 1.4$ for the best-fit three-planet model and the rms $\sim 4$~m/s
has $\sim 1.2$~m/s excess over $\sm \sim 2.8$~m/s (the mean of
${\sigma_{\idm{m}}}$ is $\sim 1.9$~m/s).  In such a case, the three-planet model
is not fully adequate for explaining the RV variability and that the presence of
an additional, yet undetected planet, is statistically justified. 

%
%------------------------------------------------------------------------------
%
\section{Four-planet system and its orbital stability}
%
%------------------------------------------------------------------------------
%
To verify the hypothesis about the four-planet system, we again used  the hybrid
code driven by the kinematic model of the RV. This time we assumed that {\em
all} orbital periods are in the range of $\sim[8,6500]$~days. The statistics of
the best fits  gathered in the search are illustrated in Fig.~\ref{fig:fig5}. We
marked $\sim 1000$ different solutions yielding $\Chi$ within the $3\sigma$
confidence interval of the best fit with $P_{\idm{c}}\sim 4000$~days,
$\Chi=0.626$,  and an rms of 2.276~m/s (marked by crossing lines). However,
after a check by MEGNO integrations \citep{Cincotta2000}, we found that this fit
leads to a chaotic configuration, so we selected another stable
(quasi-periodic)  solution with very similar $\Chi=0.627$, an rms of 2.276~m/s
and $P_{\idm{c}}\sim 4500$~days. Its orbital parameters are given in Table~2 and
we call it Fit~II from hereafter. We use that initial condition to demonstrate
some orbital properties of the  \muarae{} system. We tried to refine this
solution by the self-consistent $N$-body model of the RV but we could not
improve it significantly; the osculating elements also did not change.

In general, the four-planet solutions have much smaller rms' (by $\sim
1.7$~m/s)  than the three-body configurations and they impose the new, $\sim
0.5~\mbox{m}_{\idm{J}}$-mass Jovian planet, much closer to the star than the
companion \hd{}b in $\sim 650$~days orbit detected a few years ago. The most
striking feature of the four-planet configurations is low eccentricity of all 
orbits. They are roughly less than 0.2 for all Jovian planets. The innermost
hot-Neptune planet~d has the initial $e_{\idm{d}} \sim 0.2$, nevertheless, it is
not well constrained and any value in the (assumed) range [0,0.3] is equally
likely in terms of the $1\sigma$ confidence interval. Simultaneously, the
orbital periods of the two inner Jupiter-like companions, $P_{\idm{b}}\sim
646$~days and $P_{\idm{e}} \sim 307$~d appear to be bounded very well with the
accuracy range of a few days. This is not the case for the outermost planet~c
--- the $1\sigma$ error of $P_{\idm{c}}$ is about 1500~days.  Thus the
four-planet model changes completely the topology of $\Chi$.  Yet the outermost
planet would have the semi-major axis very similar to that of Jupiter. According
to the conclusions of \cite{Santos2004}, the orbit of planet~d  should be almost
edge-on. Hence assuming that the entire system is coplanar, we may expect that
the minimal masses determined from the Keplerian fits are likely close to the
real ones.

Figure~\ref{fig:fig6} illustrates in subsequent panels the synthetic curve of
the best-fit system (Fit II, Table~2) and the period-phased RV signals of the
planets d, e, b, and~c. The resulting curve closely follows all the measurement
points. The rms is only $\sim 2.3$~m/s.  As we expected, its value is comparable
with the joint error $\sigma$ if we assume that $\sigma_{\idm{j}}  \sim 2$~m/s.
This indicates a statistically perfect solution.   The next panel shows the raw
Lomb-Scargle periodogram \citep{Press1992} of the data.  Besides the dominant
signal of planet~b, there are only visible  peaks about  32~days and 225~days.
Simultaneously, the periods of $\sim 9$~days and  $\sim 307$~days seem to be
completely absent in the periodogram. The period of $\sim 32$~days is close to
the rotational period  $P_{\idm{rot}}$ of the star derived from the index $\log
R'_{\idm{HK}} = -5.034$  by \cite{Santos2004}. However, it is not consistent
with another estimate of $\sim 22$~days by \cite{Bouchy2005} who analyzed the
short-term variability of the RV spectrum of \muarae{}. We have no good 
explanation for the $\sim 225$ days period  as we did not find any  reasonable
solution consistent with its value.  Most likely, it is an alias of the $\sim
32$~days period.  Instead, there exist other apparently precise fits having
periods uncorrelated with the periodogram,   
for instance, 
(2.332~m/s, 90.19~days, 0.30, 3.739~deg,
14391.686~days), (12.298~m/s,    517.578~days,      0.581,    157.537~deg, 
13841.816~days), (39.010~m/s,    621.323~days,      0.267,    259.890~deg, 
10908.948~days), and (21.580~m/s,   2459.212~days,  0.475,    183.853~days,
13542.752~days), in terms of parameter tuples of Eq.~1, yielding an rms $\sim
2.7$~m/s ($T_{\idm{p}}$ are shifted by JD2,440,000). However, these solutions
are very unstable. We analyse such alternative solutions in Section~5.1.

Another unusual coincidence of periods is the ratio $P_{\idm{e}}/P_{\idm{d}}
\sim 32$. That may also indicate an aliasing in the data. Nevertheless, the
signal of $K_{\idm{e}} \sim 14$~m/s exceeds by a few times the errors of data 
and our experiments show that it cannot be fitted well by the signal of the
hot-Neptune planet itself.  Still, it remains possible that the periodic signal
of the putative planet~e is an alias of the rotational period because
$P_{\idm{e}} \sim 10 P_{\idm{rot}}$ if $P_{\idm{rot}} \sim 32$~days and
$P_{\idm{e}} \sim 14 P_{\idm{rot}}$ if $P_{\idm{rot}} \sim 22$~days. Then the RV
variability could be attributed to  a moving spot on the star surface. However,
in the observations of \cite{Santos2004} there is no sign of abnormally large
variations of the RV (exceeding the amplitude of the hot-Neptune signal)  during
about 80~days, covering already a few $P_{\idm{rot}}$.  It is a good argument
justifying the planetary origin of the signal with the period of $\sim
307$~days.

Finally, we investigate whether the hot-Neptune \hd{}d can be reliably detected
in the AAT data alone. Because we fit a multi-period orbital model to a  small
number of measurements, there is always a risk of generating a periodic signal
through random fluctuations. To determine the confidence level of a weak signal
in the data, we  apply the so called test of scrambled residuals
\citep{Butler2004}, see also our paper \cite{Gozdziewski2006b}. Having the best
fit (Fit~II) in hand, we remove the synthetic RV contributions of the Jovian
planets from the  measurements. Next, we randomly scramble the residuals,
keeping the exact moments of observations, and we search for the best-fit
elements of the putative planetary signal.  To speed up the search, we limit the
period range to [2,128]~days. We use again the hybrid code. Thanks to a large
population size (1024) and the number of generations (256), the GAs  reliably
find the best fits to the single-planet model; yet the fits are then refined
with the simplex.   After many repetitions of such a procedure, we get a
Gaussian-like  histogram of $\Chi$ of the best fits to the synthetic sets of
residuals. If the real data were uncorrelated, the value of their $\Chi$ should
be found in the range spanned by the histogram. That histogram of $\sim 36,000$ 
Keplerian fits to scrambled residuals is shown in the right-bottom panel of
Fig.~\ref{fig:fig6}.  Clearly, the likelihood that the residual signal is only a
white noise is negligible. For a reference, we computed the Lomb-Scargle
periodogram of the original residuals (the left-bottom panel in
Fig.~\ref{fig:fig6}). The strongest peak at $\sim 9.637$~days is consistent with
the estimate of $P_{\idm{d}}$ in the hybrid search (see Fig.~\ref{fig:fig5}).  
These results support independently the detection of the hot-Neptune  \hd{}d by
\cite{Santos2004} and prove the excellent quality of the AAT data.
%
%------------------------------------------------------------------------------
%
\subsection{Stability analysis}
%
%------------------------------------------------------------------------------
%
To investigate the long-term stability  of the best-fit configuration, we
employed the MEGNO indicator computed by the symplectic algorithm
\citep{Gozdziewski2005b}. In the first test, we integrated the entire
four-planet system (with the initial condition Fit~II  given in Table~2) over
$10^5$~yr.  It appears to be rigorously stable which according to the theory of
the fast indicator is indicated by MEGNO rapidly converging to the value of 2.
Because  the low-mass innermost planet revolves very close to the parent star, 
its influence on the secular dynamics is negligible. Thus in  the next
experiments we considered the three-Jovian system only. The results of the
integrations conducted over 50~Myr are shown in Fig.~\ref{fig:fig7}.  Such
period of time corresponds to $\sim 5 \cdot 10^6 P_{\idm{c}}$ and is long enough
to detect possible secular instability. Still the MEGNO converges perfectly to~2
and that indicates a stable quasi-periodic solution. Indeed, the eccentricities
and semi-major axes are almost constant over this time.

In order to illustrate the dynamical environment of the \muarae{} system, we 
also computed the dynamical maps in terms of the Spectral Number  and $\max e$
in the ($a_{\idm{b}},e_{\idm{b}})$-plane (Fig.~\ref{fig:fig8}). These maps
reveal that the nominal position of planet~b is close to the island of 2:1~MMR  
with the planet~e. Yet none of the critical angles of this resonance librate in
the nominal best-fit configuration. We found only some signs of the apsidal
anti-corotation (see the left-lower panel in Fig.~\ref{fig:fig7}). Such
dynamical character of the \muarae{} system is puzzling in the light of what is
known about  four extrasolar systems presumably locked in  the exact 2:1~MMR,
i.e., Gliese~876 \citep{Rivera2005}, HD~82943 \citep{Mayor2004,Lee2006}, 
HD~128311 \citep{Vogt2005} and HD~73526 \citep{Tinney2006}. The \muarae{} system
seems only close to the border of this resonance. A possible explanation of this
behavior is given in the next section.

One should be aware that $K_{\idm{c}}$  (and the resulting minimal masses) as
well as $P_{\idm{c}}$ can be significantly varied within the $1\sigma$
confidence range of the best fit. Moreover, not all Keplerian fits shown in
Fig.~\ref{fig:fig5} are necessarily stable. Thus the GAMP search  would be very
helpful to get a more detailed self-consistent statistics of the stable
solutions.  In this work we skipped such a test due to its significant numerical
expense caused by the extremely different orbital periods.  Instead, we carried
out a more direct check of the stability preserving the  full accuracy of the
four-planet fits.  First,  the $\sim 1000$ different best Keplerian fits
gathered in the hybrid search were refined by the Newtonian model of the RV.
Next, we performed long-term MEGNO integrations of these self-consistent
solutions.  At this stage, the planet~d has been skipped.  The integration time
$t_m \sim 3\cdot 10^5 P_{\idm{c}}$   ($\sim 3$~Myr) is long enough to detect
unstable solutions related to the short-term two- and three-body MMRs and also
strongest secular resonances.  A detection of all secular instabilities would
require much longer integration times $\sim 10^7$--$10^8$~yr because the secular
periods, as the apsidal period of the outermost orbit, are $\sim 10^5$~yr. 

The results of this experiment are shown in the top row of Fig.~\ref{fig:fig9}.
The best-fit solutions have an rms $\sim 2.27$~m/s. Globally there is only a
little improvement of the Newtonian fits when  compared to the Keplerian
solutions. The mutual interactions are not yet clearly evident in the AAT data. 
The overall distribution  of the Newtonian fits in the
$(a_{\idm{c}},e_{\idm{c}})$-plane mimics the projection of the kinematic
solutions onto the  $(P_{\idm{c}},e_{\idm{c}})$-plane (see
Fig.~\ref{fig:fig5}).  There is a well defined minimum of rms [and $\Chi$] in
the $(a_{\idm{e,b}},e_{\idm{e,b}})$-planes, but the semi-major axis of planet~c
may be varied over 2~AU within the limit of rms $<2.3$~m/s. We examined the
stability of all such fits. The stable  initial conditions (with
$|\Ym(t_m)-2|<0.001$) are marked with filled yellow circles. We notice that the
stable fits appear for $e_{\idm{e,b}}<0.1$. It remains likely that by altering
the orbital phases or other parameters (like masses) within the acceptable error
bounds, we can find reasonably precise and stable solutions also for
$e_{\idm{e,b}}>0.1$. However, this procedure would require the self-consistent 
GAMP-like search.

%
%------------------------------------------------------------------------------
%
\section{Four-planet model revisited}
%
%------------------------------------------------------------------------------
%
Shortly after submitting the  original manuscript, we learned about an
independent work by \cite{Pepe2006}.  Our conclusions are in a great agreement 
with the results of their analysis although these authors study a completely
different  and independent set of the RV data: the measurements published by
\cite{McCarthy2004}, data gathered with CORALIE and observations collected
during $\sim 2$~year campaign with the HARPS spectrometer.  Having access to the
new extended  observations, we can discuss some of the conclusions derived on
the basis of the new AAT data alone.

In a number of experiments, we considered three data sets:  S1 --- 108~points by
the AAT \citep{Butler2006} as analyzed already,  S2 --- the set S1 extended by
the CORALIE measurements (with errors rescaled by $\sigma_{\idm{j}}\sim
3.5$~m/s) and 78~measurements from HARPS (the original errors are unmodified) 
and S3 --- the set composed of only the AAT and HARPS data. Note, that from the
CORALIE and HARPS observations \citep{Pepe2006}, we substracted  the mean of all
RV measurements in the given set, respectively.

Using the most precise S3 set which covers the whole observational window,  we
carried out the hybrid search for the four-planet Keplerian fits including two
independent instrumental RV offsets. As in the previous test, the orbital
periods of {\it all} planets are bounded to the range of $[8,6500]$~days. The
best Keplerian fit is similar to Fit~II, i.e., in terms of the parameter tuples
of Eq.~1,
 (3.118~m/s,      9.636~days,      0.164,     211.68~deg, 302.852~days),
 (10.849~m/s,    307.937~days,      0.127,    180.752~deg, 3323.904~days),
 (37.147~m/s,    649.500~days,      0.000,    231.352~deg, 2734.35~days), 
 (24.056~m/s,   4293.506~days,      0.054,     90.829~deg, 3319.1~days),
 for planets d,e,b,c, respectively and velocity offsets
$V_0=  -10.932$~m/s, $V_1= 11.209$~m/s
($T_{\idm{p}}$ are shifted by JD2,450,000). This fit has
$\Chi\sim 1.48$ and an rms $\sim2.51$~m/s.

Next, we refined $\sim 2500$ different solutions gathered in the hybrid search
using the Newtonian code.  Similarly to the case of the AAT data only (S1), we
also performed the stability check of the best-fit configurations.  For the S2
set (see the middle panels in Fig.~\ref{fig:fig9}), the smallest rms of the
best-fit solutions is $\sim 3.66$~m/s, significantly more than that one obtained
for the S1 set alone (2.3~m/s);  likely due to a relatively low accuracy and
scatter of the CORLIE data. The best fits to the S3 data  yield an rms $\sim
2.5$--$2.7$~m/s  also larger by $\sim 0.3$-$0.4$~m/s from the rms obtained for
the S1 set. Also their $\Chi\sim 1.5$ suggest that the errors of the HARPS
measurements could be underestimated in the long-run. Hence finally we minimized
the rms instead of $\Chi$. In this case (see the bottom panels of
Fig.~\ref{fig:fig9}), the rms of the best fits is smaller than for the S1 set
alone, $\sim 2.2$~m/s. For instance, a stable solution, given in terms of tuples
$(m~[M_{\idm{J}}], a~[\mbox{AU}], e, \omega~[\mbox{deg}], {\cal M}~[\mbox{deg}])$:
(0.032, 0.09286, 0.135, 204.663, 272.050),
(0.506, 0.9393, 0.083,  204.913, 293.54),
(1.696, 1.533, 0.062,  45.56, 4.039),
(1.926, 5.134, 0.053,  103.06, 156.057), for planets d, e, b, c,
respectively, with velocity offsets
$V_0=-8.466$~m/s and $V_1=13.695$~m/s yields an rms $\sim 2.22$~m/s.

Still, for all the analyzed sets S1, S2 and S3, the orbital parameters
($a_{\idm{c}},e_{\idm{c}}$) are allowed to vary in relatively wide ranges
qualitatively similar to the ones we determined for the AAT data only. Yet the
shapes of the minima are much more irregular than those found for the set S1.
Clear limits of stable solutions in the ($a_{\idm{b,e}},e_{\idm{b,e}}$)-planes
are evident.

Having the ensemble of stable fits, we also try to explain the apparent lack of
locking of the subsystem e-b in the 2:1~MMR. Because the elements of the inner
giants are well constrained, we selected three best fits to the S3 data (all
yielding an rms $\sim 2.2$--$2.3$~m/s) with $a_{\idm{c}} \sim 4.7, 5, 6.3$~AU,
respectively; also Fit~II  ($a_{\idm{c}} \sim 5.5$~AU) may be put in this short
sequence. Next, we calculated the dynamical maps in the
$(a_{\idm{b}},e_{\idm{b}})$-plane in the neighborhood of the selected
solutions.   These maps (Fig.~\ref{fig:fig10}) reveal very sharp borders of
stable motions around $e_{\idm{b,e}}\sim 0.1$, in an excellent agreement with
the MEGNO tests (Fig.~\ref{fig:fig9}). The outermost planet strongly modifies 
the dynamics of the e-b pair. For a moderate distance of planet~c from the e-b
subsystem, there is no evident 2:1~MMR island at all. It shows up for
$a_{\idm{c}}$ larger than $5$--$5.5$~AU and  for relatively large eccentricities
(see also Fig.~\ref{fig:fig5}).  However, such large eccentricities are ruled
out  in the model by the stability constraints.
%
%------------------------------------------------------------------------------
%
\subsection{The 1e:1b~MMR hypothesis}
%
%------------------------------------------------------------------------------
%
Remarkably, the Keplerian and $N$-body fits to all the data sets (S1,S2,S3) reveal a
number of low rms solutions corresponding to the 1b:1e MMR of the planets e and~b. 
This possibility is also investigated by \cite{Pepe2006} but they did
not report any stable 1:1 MMR solutions. 

The 1:1~MMR configurations are quite frequent in the Solar system. There are
also speculations on the existence of such extrasolar configurations
\citep{Laughlin2001,Nauenberg2002,Gozdziewski2006c,Ford2006}. For a closer
analysis we selected the fits to the S3 set. The Newtonian best-fit solutions
with $a_{\idm{e}}\sim a_{\idm{b}}$ are illustrated in the top-left panel of
Fig.~\ref{fig:fig11}. Their quality is comparable with the ones related to the
2e:1b~MMR. Nevertheless, all these 1e:1b~MMR configurations are strongly
unstable leading to a quick collision between the planets.  Due to extremely
strong dynamical interactions, even small errors of the phases or other
parameters may be critical for the system stability. Thus in a close proximity
of apparently unstable solutions, stable systems consistent with the RV
observations may still exist.  
We tried to ``stabilize'' such fits with
GAMP, in terms of the three-Jovian planets model (that is again due to the
efficiency reasons).
Unfortunately,  we did not find any long-term stable  best-fits as precise as
the ones of the 2e:1b~MMR configurations. However, there exist {\em stable}
1e:1b~MMR solutions with reasonably small rms $\sim 5$~m/s
($\sim 4.5$~m/s for the whole, four-planet system).
A synthetic curve of
an example configuration fitting well the most precise AAT and HARPS
measurements, is illustrated in the top-right panel of  Fig.~\ref{fig:fig11}
(the orbital elements are given in the caption).  A striking result is
illustrated  in the dynamical map of that best-fit (see the right-bottom panel
of Fig.~\ref{fig:fig11}).  It reveals the island  of stable motions
extending over 0.2~AU in $a_{\idm{e}}$ and covering {\em the entire} range of
$e_{\idm{e}}$! The selected solution is stable over the secular time-scale.  We
show the results of 1~Gyr direct integration of the system including the planets b,c
and~e (see the left-bottom  panel in Fig.~\ref{fig:fig11}).  The apsides of  e-b
subsystem are anti-aligned, librating around $180^{\circ}$  with the
semi-amplitude of $\sim 70^{\circ}$ and with  $e_{\idm{b,e}}$ varying up to 0.5
and $e_{\idm{c}}\sim 0.1$.  We also found other solutions with the apsides
librating about centers different from $180^{\circ}$.  The 1e:1b~MMR island has
a very complex dynamical structure, particularly in the vicinity of the best
fit.  Direct integrations show that weakly chaotic solutions in that zone may
lead to a sudden collision of planets after hundreds Myr of apparently stable
orbital behavior. Let us note that to refine the stable fit, the MEGNO was 
integrated over $\sim 10^5~P_{\idm{c}}$.

Still, there is an open question whether the fit quality should be  the final
argument to rule out the 1b:1e~MMR configurations.  We think that the problem
deserves a further study. Let us recall that we searched only for coplanar
solutions. The stability may be easily preserved for inclined configurations
\citep{Gozdziewski2006c}. Yet the system architecture involving planets b and~e
in the 1:1~MMR may still permit the existence of other bodies besides the well
established planet~c.

%
%------------------------------------------------------------------------------
%
\section{Conclusions}
%
%______________________________________________________________________________
%
The long-term radial velocity surveys of the Sun-like stars constantly reveal
more and more exciting features of the planetary systems.    The \muarae{}
system may be the second known four-planet configuration, after 55~Cnc
\citep{McArthur2005}.  Remarkably, in such a multi-planet  system the orbits are
close to circular ones, similarly to three-planet systems of HD~37124
\citep{Vogt2005} and HD~68930 \citep{Lovis2006} resembling the Solar system
architecture. The alternative best-fit three-planet configurations  may contain
two Jupiter-like planets in the 4:1~MMR.  In that case, the eccentric orbits
($e_{\idm{b,c}}\sim 0.3)$ would be localized in a dynamically active region of
the phase space; in fact, on the edge of dynamical stability. Besides the worse
fit quality, it could be a heuristic argument against the three-planet system.
Obviously, the key for the proper understanding of the orbital architecture is
the improved precision of observations and their extended time span. Curiously, 
the new data reveal a Jovian planet  that has the orbital period two times
shorter than the companion already detected  a few years ago. It remains an open
question whether the two inner companions (the planets b and~e) in the
four-planet configuration are locked in the 2:1~MMR. Most likely, the presence
of the massive companion~c prevents the creation of the 2:1~MMRs island known in
the other four extrasolar systems  (Gliese~876, HD~82943, HD~128311, and
HD~73526) presumably locked in this resonance. In any case, even the proximity
of their orbits to such particular dynamical state  can be counted as the fifth
occurrence of the 2:1~MMR among $\sim 20$ multi-planet systems known up to date.
It may indicate a universal dynamical mechanism governing the creation and
orbital evolution of extrasolar planetary systems.  We also found stable
configurations related to  the 1:1~MMR of the inner Jovian planets in eccentric
orbits. However, such fits are significantly worse than those derived for the
system with quasi-circular orbits. The results of our analysis permit us to
conclude that a few years of observations are still required to constrain the
outermost orbit without a doubt.   Looking at the orbits of the \muarae{}
planets  and recalling the results of \cite{Laskar2000}, we see a free space in
the range of $\sim (0.2,0.8)$~AU in which  new planetary objects may yet exist. 
Extensive numerical simulations concerning the less constrained parameters of
the outermost planet would be necessary to answer this question. 
%
%------------------------------------------------------------------------------
%
%\section{Acknowledgments}
%
%------------------------------------------------------------------------------
%
\acknowledgements
We would like to strongly  acknowledge the Anglo-Australian Telescope Team for
publishing the precision RV measurements. This work could not be done  without
the access to these data. We thank the anonymous referee for the review that
helped us to clarify and improve the paper. We kindly thank Zbroja for the
correction of the manuscript. This work is supported by the Polish Ministry of
Sciences and Education,  Grant No.~1P03D-021-29 and by the N. Copernicus
University, Grant No.~367-A. 

%
%------------------------------------------------------------------------------
%

%
%------------------------------------------------------------------------------
%
% T A B L E S
%------------------------------------------------------------------------------
%
% TABLE 1
%
\begin{table}
\label{tab:tab1}
\caption{
Primary parameters ($K,P,e,\omega,T_{\idm{p}})$ of the three-planet Keplerian
model (in Eq.~1) and the inferred {\it astrocentric} orbital elements of the
best-fit~I found in this paper for the RV of \muarae{} \citep{Butler2006}.  The
jitter estimate  $\sigma{\idm{j}}=3.5$~m/s, the mean measurement errors
$\sm=1.87$~m/s, and the mean combined data error 
$(\sigma{\idm{j}}^2+{\sm}^2)^{1/2}=4.02$~m/s.  Numbers in parentheses are for
the $1\sigma$ errors estimated on the basis of the best fit statistics gathered
in the hybrid search (see the text and Fig.~\ref{fig:fig3}). By definition,
$T_{\idm{p}}$ and $\omega$ are not well constrained when $e$ is small. Instead, 
the error of the orbital phase, or the  mean longitude, $\lambda(t_0) \equiv
{\cal M}(t_0)+\omega$, where  ${\cal M}(t_0)$ is the mean anomaly, is given. The
epoch of the first observation $t_0 \equiv$ JD~2,451,118.89. The reference epoch
$T_0$ is JD~2,440,000. Mass of the parent star $M_{\star}=1.15$~$M_{\sun}$
\citep{Butler2006}.
}
\centering
\begin{tabular}{lccc}
   &        &          &             \\
\hline
   &        &          &               \\
{\bf Best Fit~I} \hspace{1em} 
& \qquad \hd{}{\bf b} \qquad & 
  \qquad \hd{}{\bf c} \qquad & 
  \qquad \hd{}{\bf d} \qquad          \\
   &        &          &                 \\
\hline
$P$ [days]              &   632.013 (6)   &   2544.47 (60)  &  9.6369 (0.005) 
\\
$K$ [m/s]               &   37.97 (1)  &  17.67 (1)   &   3.03 (0.50) 
\\
$e$                     &   0.260 (0.07)   &   0.471 (0.05)   & 0.236 (0.2)  
\\
$\omega$ [deg] 		&  259.46  &  183.63 &  304.20
\\
$T_{\idm{p}}$ [JD-$T_0$]
			&   12137.86  &   13536.77 & 10545.71
\\
$\lambda(t_0)$ [deg] 	&  39.05 (7)  & 201.54 (7) &  115.54 (9) 
\\			
$m \sin i$ [m$_{\idm{J}}$]  
			&   1.70   &    1.15   &  0.034 \\
$a$ [AU] 		&   1.51   &    3.80   &  0.09285 \\
$\Chi$  		& \multicolumn{3}{c}{1.01}  \\
rms~[m/s] 		& \multicolumn{3}{c}{3.98} \\
$V_0$ [m/s] 		& \multicolumn{3}{c}{-0.638 (0.8)}  \\
\hline
\end{tabular}
\end{table}

%
% TABLE 2
%
{\normalsize
\begin{table}
\label{tab:tab2}
\caption{
Primary parameters ($K,P,e,\omega,T_{\idm{p}})$ of the four-planet Keplerian
best fit~II (Eq.~1) and the inferred {\it astrocentric} orbital elements.  This
fit is dynamically stable.  Numbers in parentheses are for the $1\sigma$ errors
estimated on the basis of the best fit statistics gathered in the hybrid search
(Fig.~\ref{fig:fig5}). See caption to Table~1, Fig.~\ref{fig:fig5} and the text
for more details.
}
\centering
\begin{tabular}{lcccc}
   &        &          &           &           \\
\hline
   &        &          &           &           \\
{\bf Best Fit~II} \hspace{1em} 
& \qquad \hd{}{\bf b} \qquad & 
  \qquad \hd{}{\bf c} \qquad & 
  \qquad \hd{}{\bf d} \qquad & 
  \qquad \hd{}{\bf e} \qquad 
\\
   &        &          &           &           \\
\hline
$P$ [days]       
         &   646.485 (1.5)   & 4472.967 (1300) &    9.6369 (0.005) & 307.475  (1.5)			
\\
$K$ [m/s]    
	 &   35.871 (1)  & 27.178  (10)  &  2.826  (1)   & 13.195 (2)
\\
$e$        
	 &   0.0001 (0.05)   &   0.027 (0.12) &   0.184 (0.2)  & 0.079 (0.06)
\\
$\omega$ [deg] 	
	 &  223.003    & 154.065  &   314.050  &   252.624
\\
$T_{\idm{p}}$ [JD-$T_0$]
	 & 12721.839   &  14171.256 & 10632.575  & 13070.393
\\
$\lambda(t_0)$ [deg] 	
         &  50.386 (2) &   268.400 (36) & 121.225 (10)   &  127.752 (10)
\\			
$m \sin i$ [m$_{\idm{J}}$]  
	 &   1.677   &    2.423   &  0.032     &  0.480        
\\
$a$ [AU] 		
	 &   1.535   &    5.543   &  0.09286     &  0.934     
  \\
$\Chi$  		& 
		 	  \multicolumn{4}{c}{0.627} 
\\
rms~[m/s] 		& 
			  \multicolumn{4}{c}{2.276} 
\\
$V_0$ [m/s] 		& 
			  \multicolumn{4}{c}{ -13.069 (10)} \\
\hline
\end{tabular}
\end{table}
}

%\clearpage

%
%------------------------------------------------------------------------------
%
% FIGURE CAPTIONS
%
%------------------------------------------------------------------------------
%

\figcaption[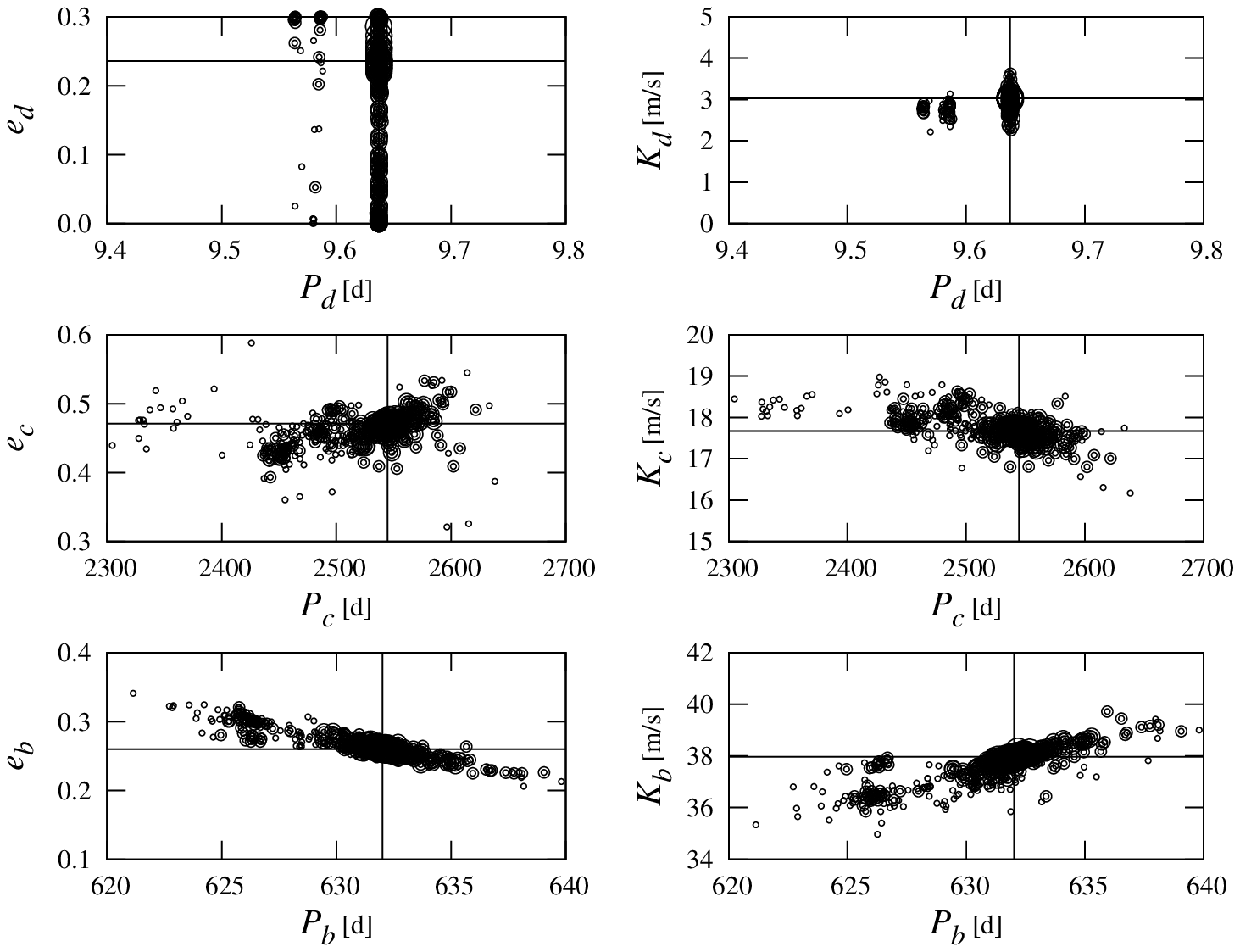]{\normalsize
   The parameters of the best-fit solutions to the
   three-planet Keplerian model of the RV of
   \muarae{}, projected onto the ($P,e$)- and ($P,K$)-plane, gathered
   in the hybrid search.
   Jitter of $\sim 3.5$~m/s is added 
   in quadrature to the measurements errors.
   The values of  $\Chi$ of the best-fit
   solutions are marked by the size of symbols.
   The largest circle is for $\Chi$ equal to 1.01;
   smaller symbols are for 1$\sigma$ solutions
   with $\Chi <1.02$, 
   2$\sigma$ solutions
   with $\Chi <1.04$, and
   3$\sigma$ solutions
   with $\Chi <1.07$ (smallest circles), respectively.
   The elements of the best fit 
   found in the search are marked by two crossing lines.
}

\figcaption[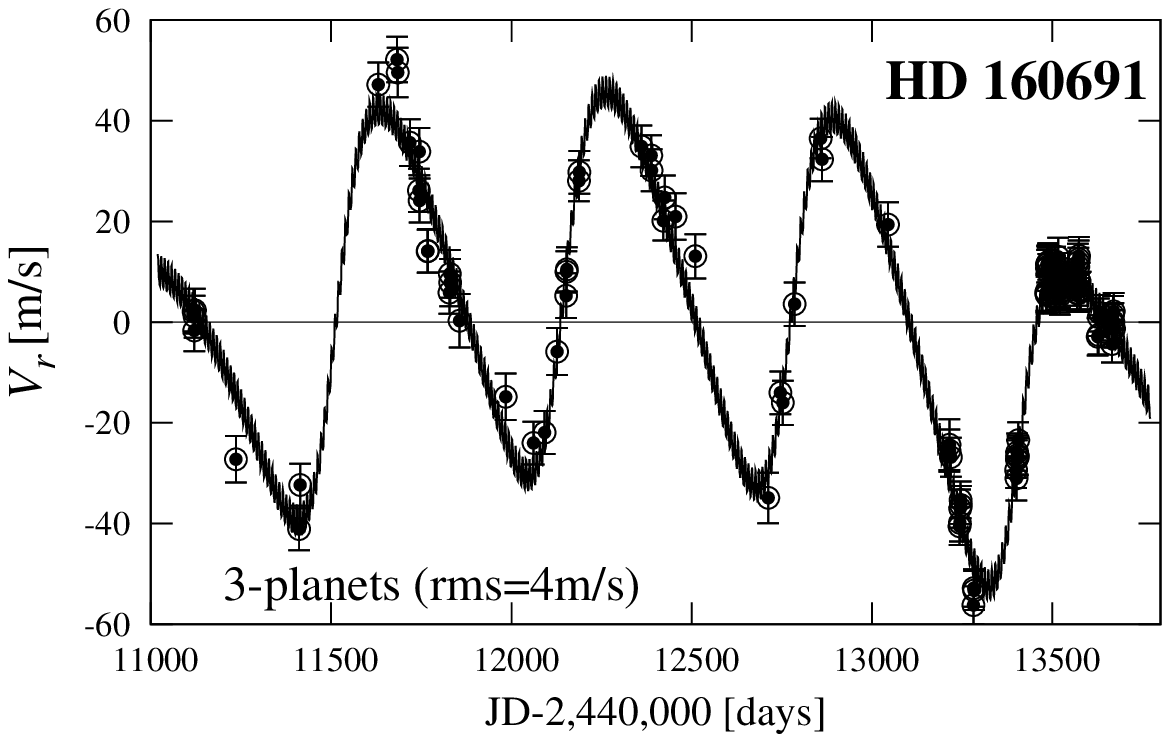]{\normalsize
The synthetic RV curve of the best-fit  Keplerian three-planet solution (Fit~I,
Table~1). The open circles are for the RV    of \muarae{} from
\cite{Butler2006}. The error bars include the measurement errors added in
quadrature to stellar jitter of 3.5~m/s.
}   

\figcaption[f3a.eps,f3b.eps]{\normalsize
The dynamical maps in the $(a_{\idm{c}},e_{\idm{c}})$-plane for the three-planet
Keplerian model of the \muarae{} system. The  large crossed circle marks the
parameters of the best Fit~I. The left panel is for the Spectral  Number, $\log
SN$. Colors used in the $\log SN$ map classify the orbits --- black indicates
quasi-periodic regular configurations while yellow strongly chaotic ones. The
right panel marked with $\max e_{\idm{e}}$ is for  the maximal eccentricity  of
planet~e attained during the integration of the system.  The thin line marks the
collision curve for planets b and~c,   as determined by $a_{\idm{b}}
(1+e_{\idm{b}}) = a_{\idm{c}} (1-e_{\idm{c}})$. The low-order MMRs of planets b
and c are labeled.  The integrations are conducted over $\sim 10^4 P_{\idm{c}}$.
The resolution is $400\times120$ data points.
}
   
\figcaption[f4a.eps,f4b.eps]{\normalsize
The dynamical maps in the $(a_{\idm{c}},e_{\idm{c}})$-plane and  the parameters
of the best-fits derived by GAMP. The edge-on, two-planet model is assumed. The
nominal initial  condition is marked by a large crossed circle. Its osculating
elements  ($m_p,a,e,\omega,{\cal M}$) at the epoch of the first observation are:
(1.572~$m_{\idm{J}}$, 1.514~AU, 0.251, 253.222~deg,  145.937~deg) and
(1.182~$m_{\idm{J}}$, 3.858~AU, 0.332, 189.385~deg,  17.771~deg), for the inner
and outer planet, respectively, and $V_0 =-0.85$~m/s. This solution yields
$\Chi\sim 1.17$ and an rms $\sim 4.7$~m/s.  The relevant parameters of other
fits within the formal $1\sigma$ confidence interval [$\Chi<1.18$ and an rms
$4.8$~m/s] of  the initial condition are marked by small circles. See the
caption to Fig.~\ref{fig:fig3} for an additional explanation.
}

\figcaption[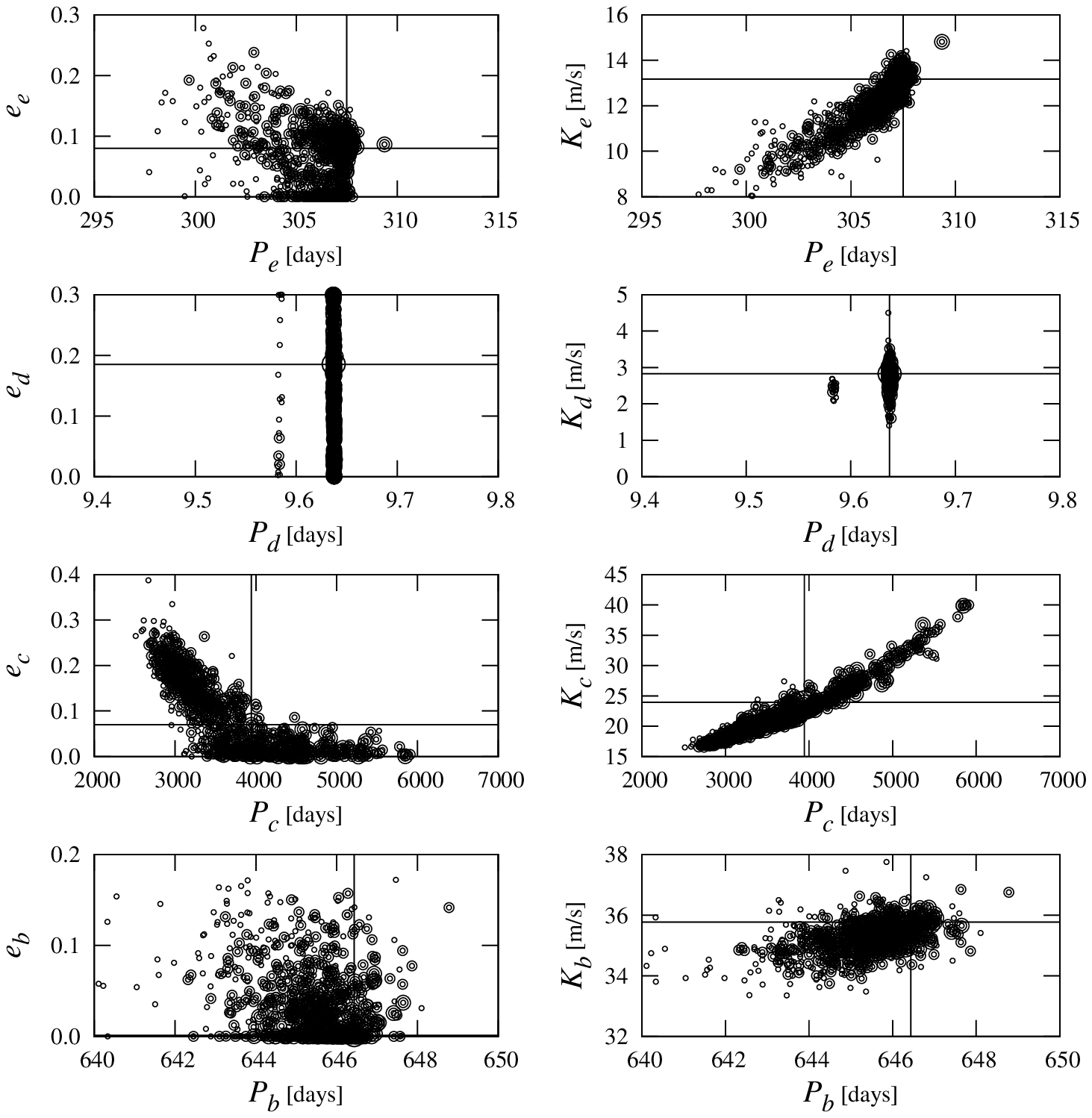]{\normalsize
   The parameters of the best-fit solutions to the
   four-planet Keplerian model of the RV of
   \muarae{}, projected onto the ($P,e$)- and ($P,K$)-plane.
    The values of  $\Chi$ of the best-fit
   solutions are marked by the size of symbols (smaller $\Chi2$---larger
circles).
   The largest circle is for $\Chi$
   equal to 0.626; smaller symbols are for 1$\sigma$ solutions
   with $\Chi <0.645$,
   2$\sigma$ solutions
   with $\Chi <0.68$, and
   3$\sigma$ solutions
   with $\Chi <0.73$ (smallest circles), respectively.
   The elements of the best fit         
   found in the search are marked by two crossing lines.
}

\figcaption[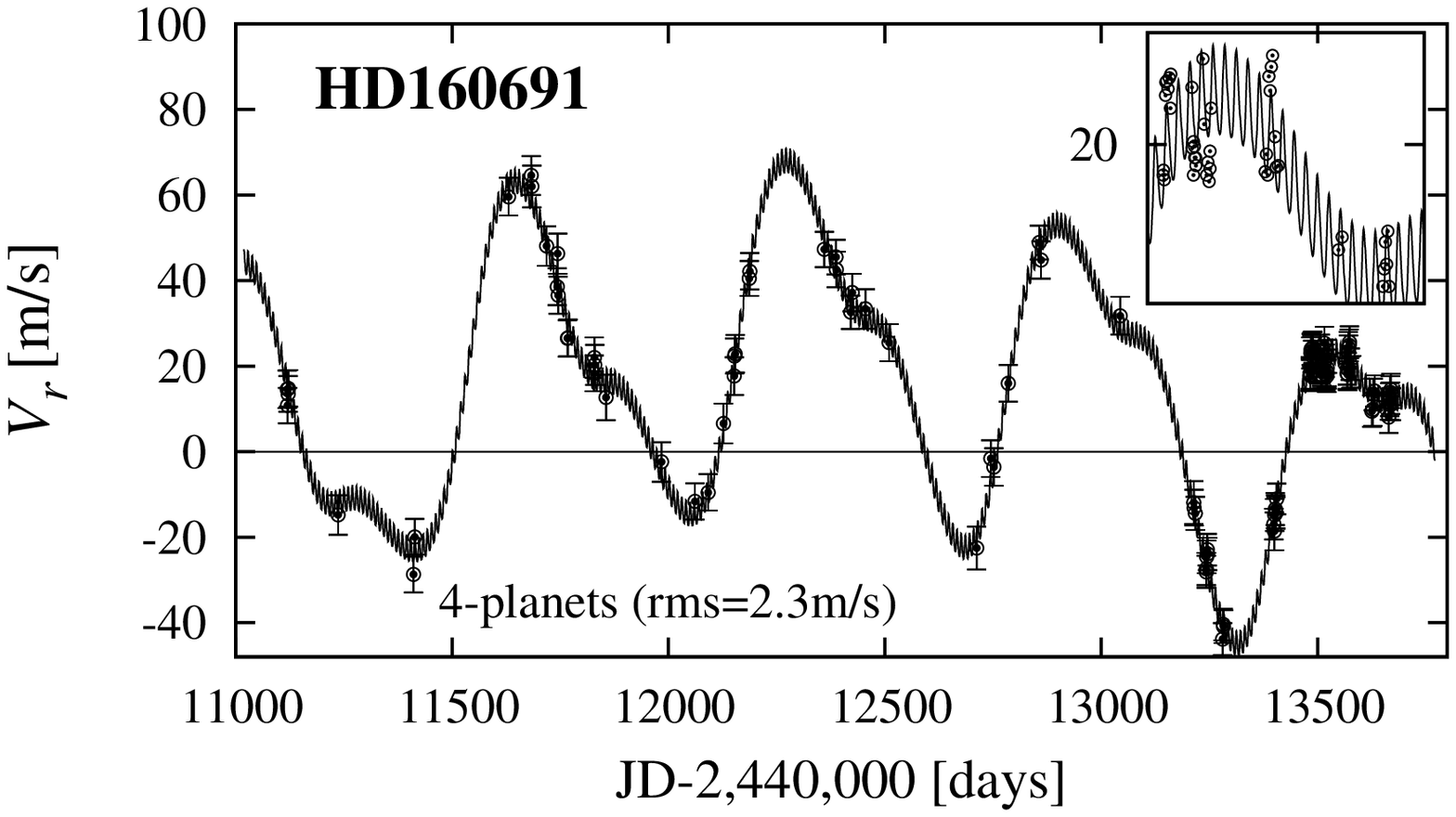,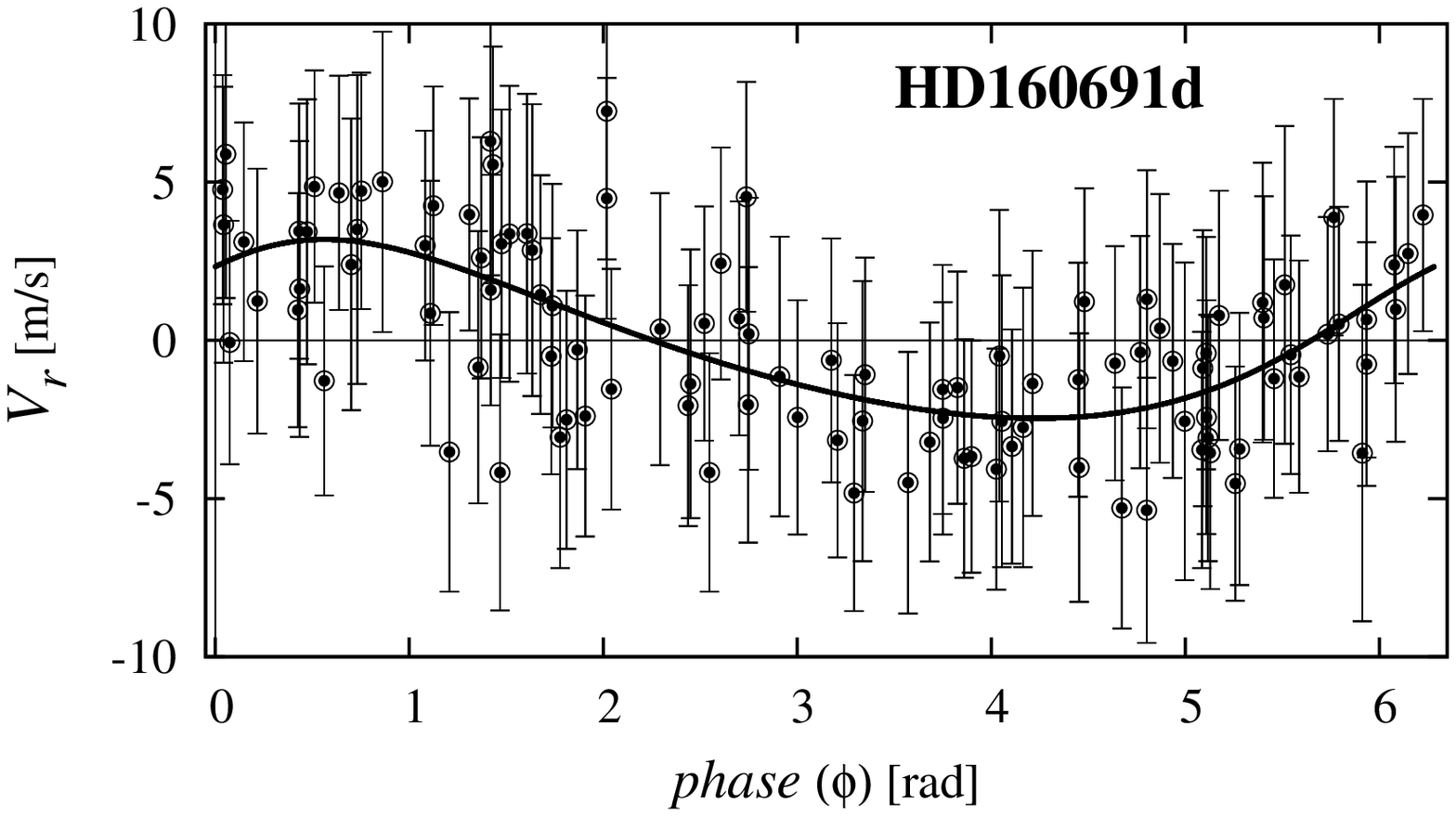,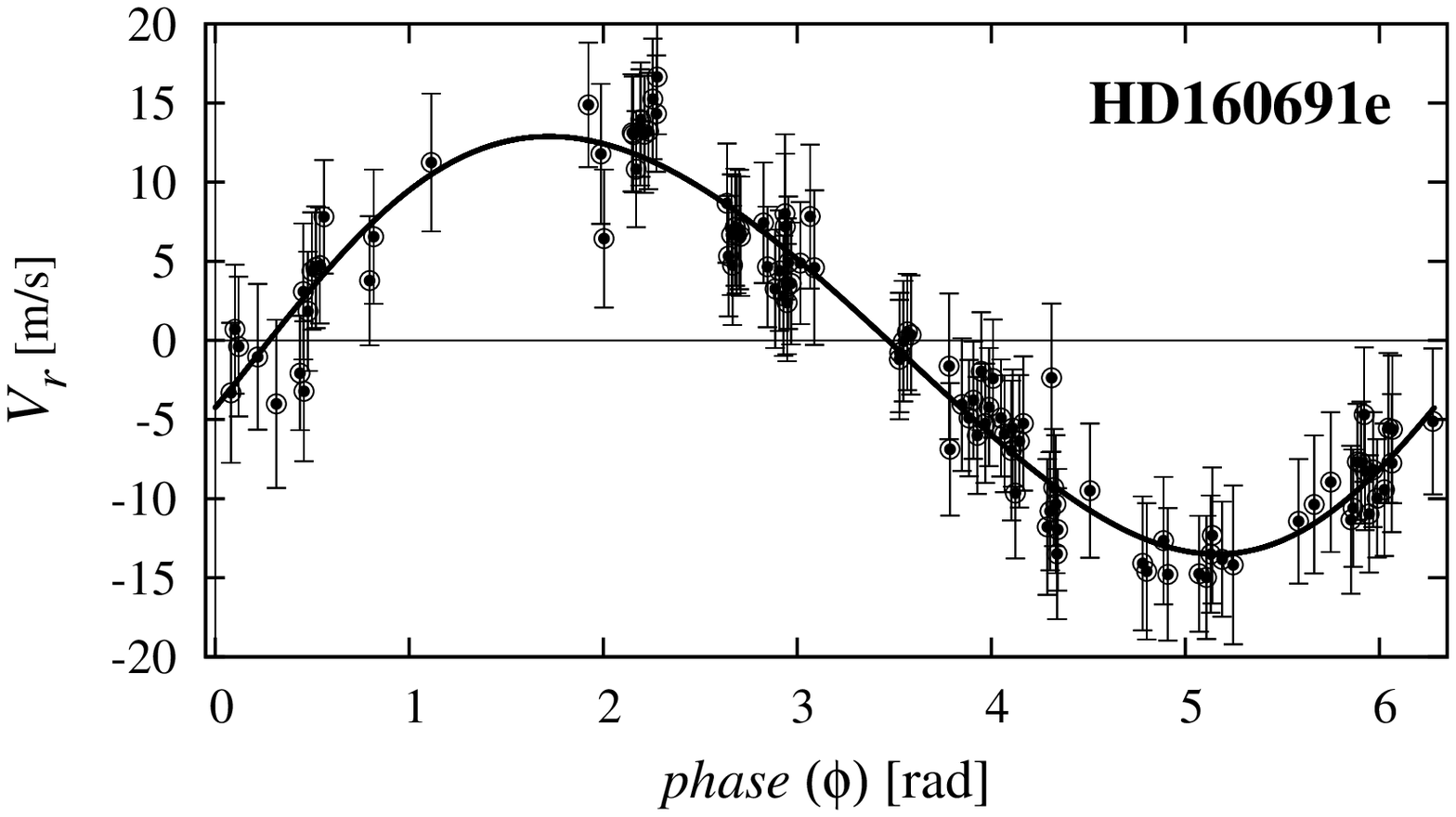,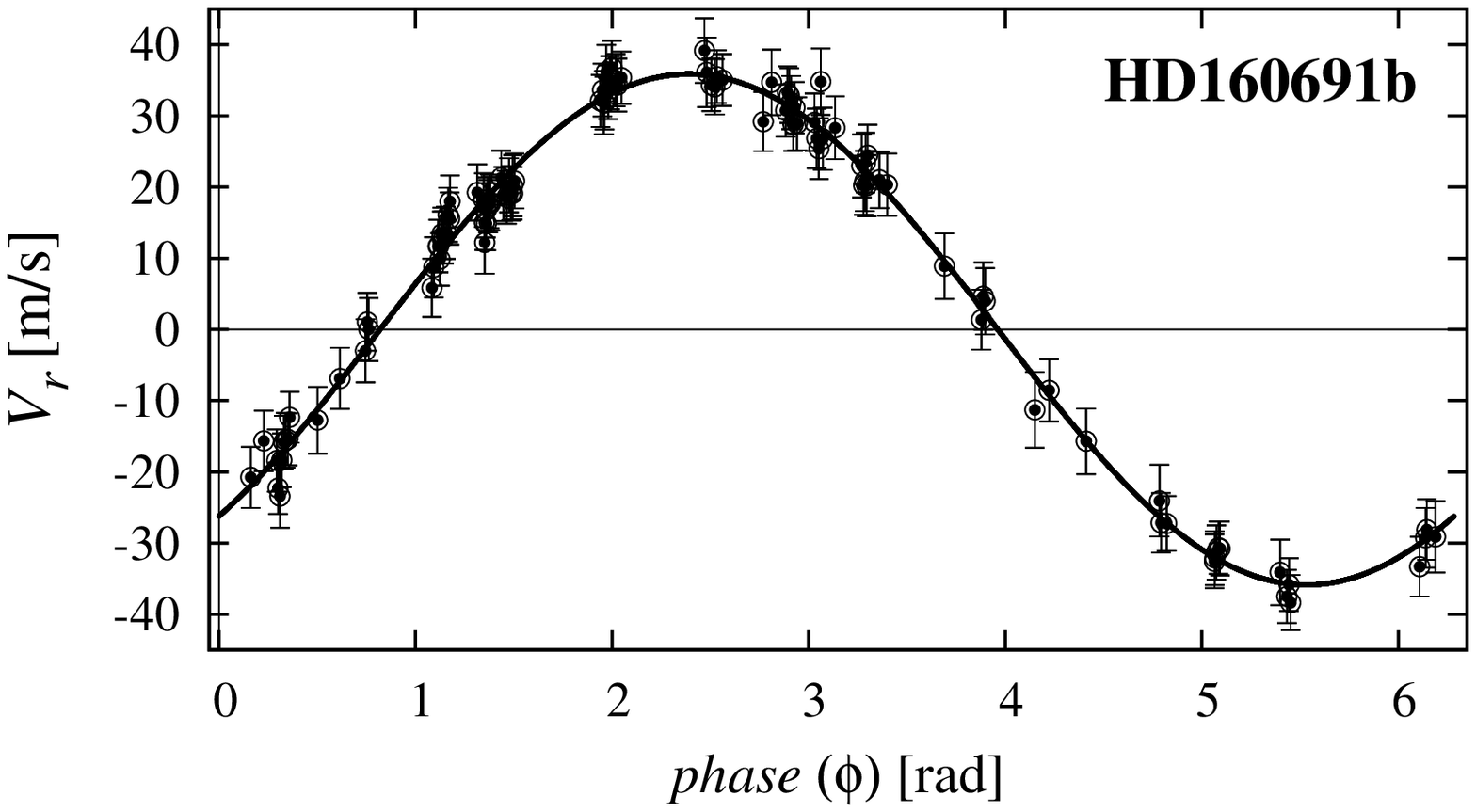,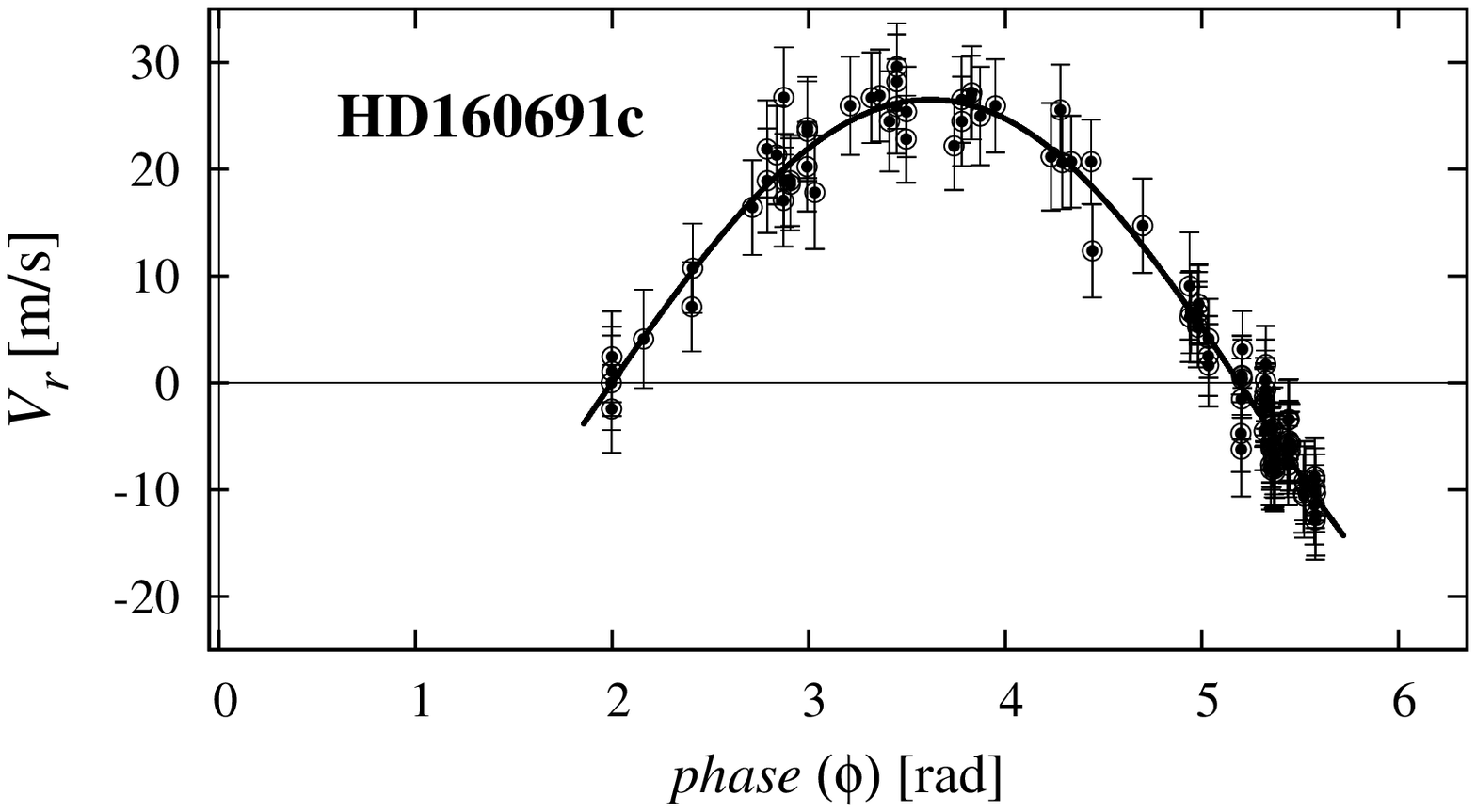,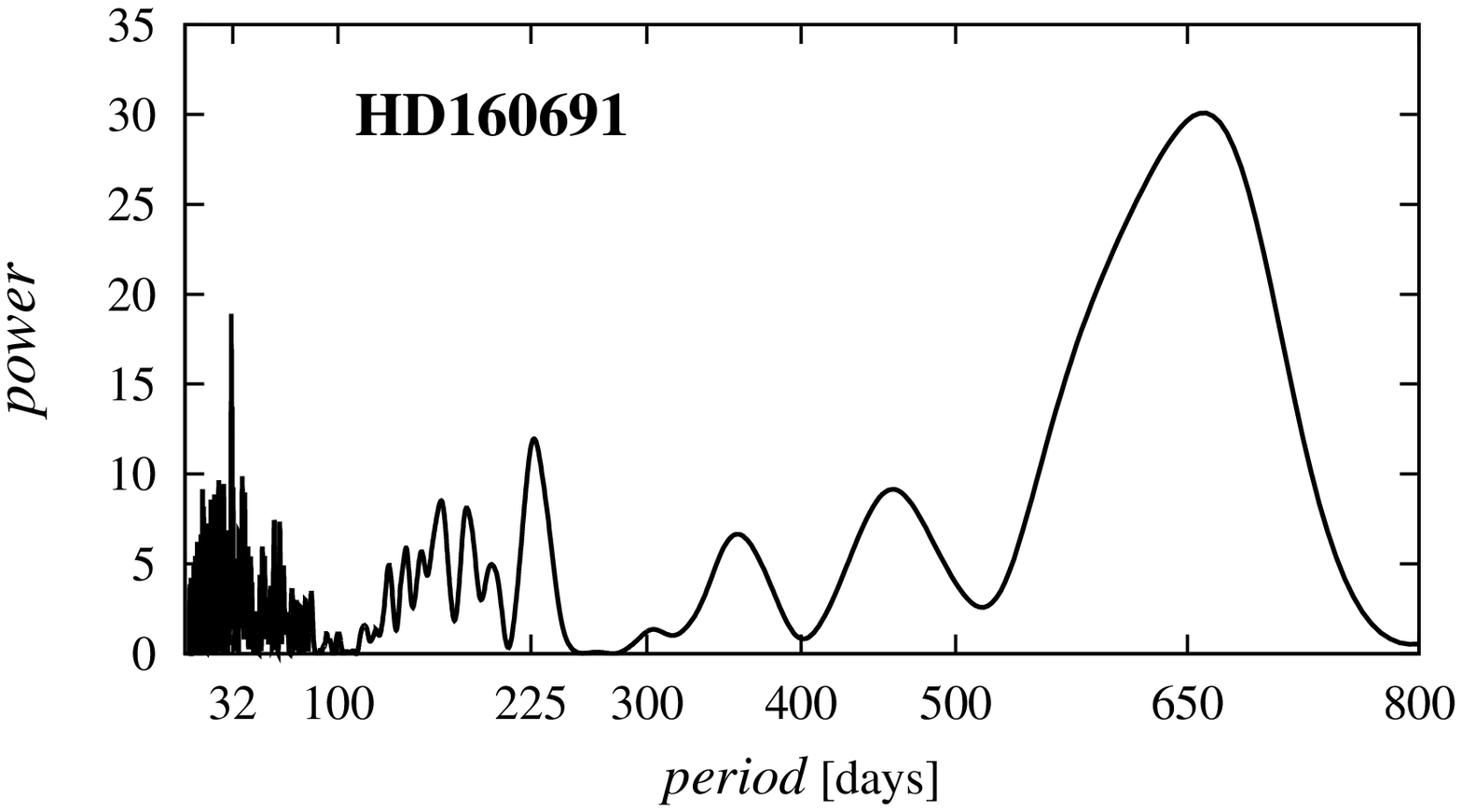,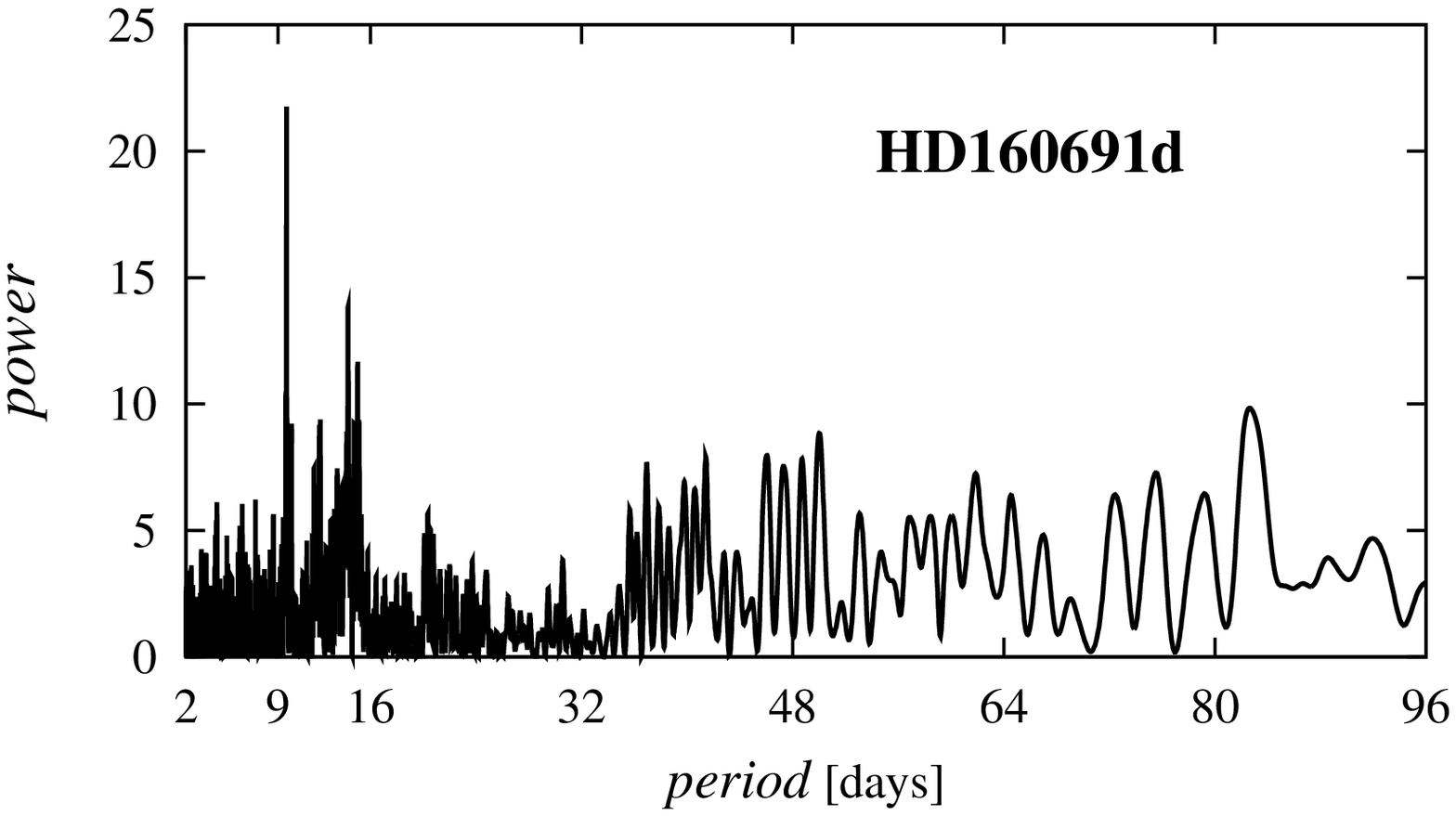,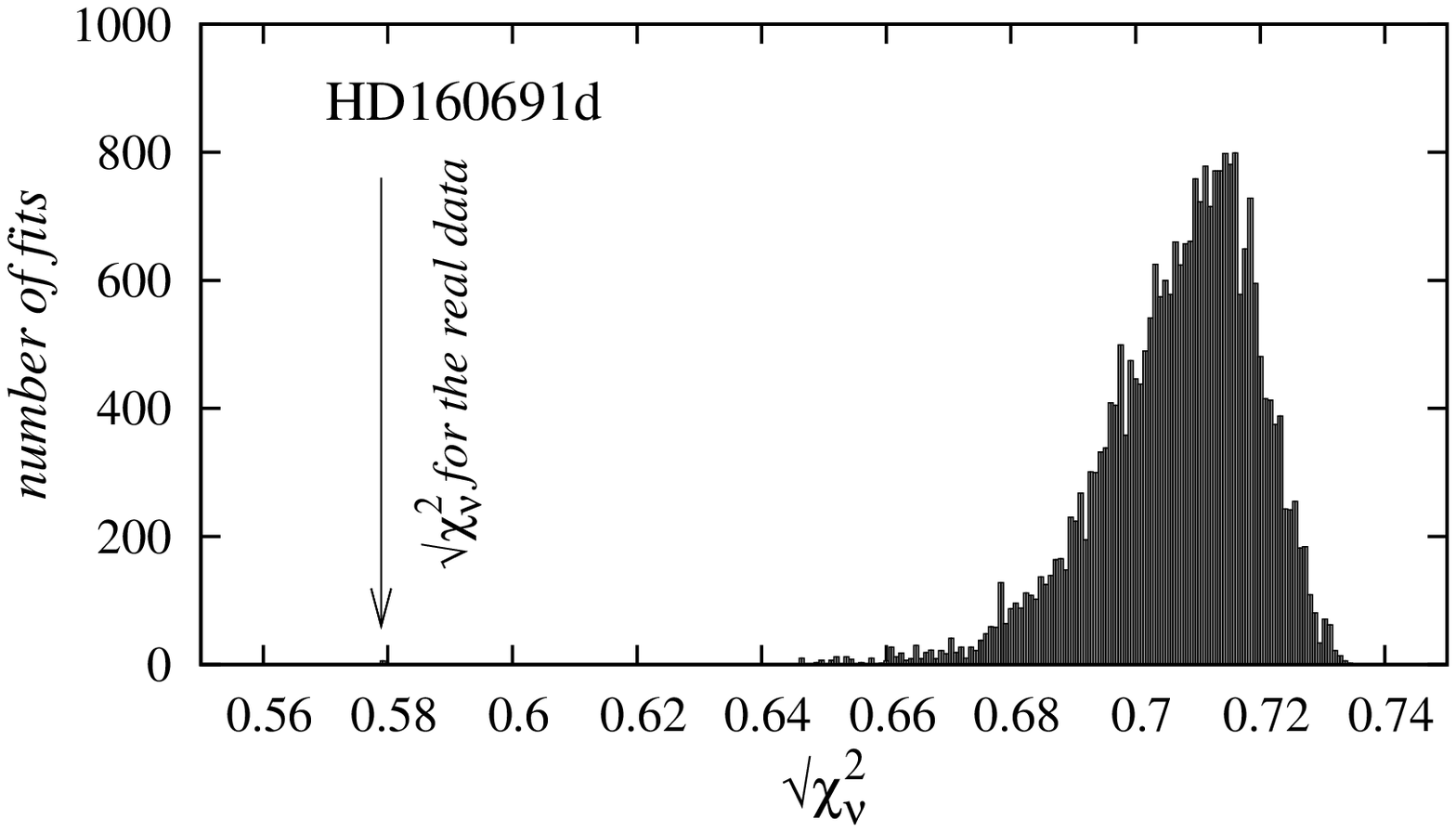]{
\normalsize
The  synthetic signal of the four-planet Keplerian best Fit~II, see Table~2. 
Subsequent panels (from the top) are for the synthetic RV signal of the
four-planet model,  the period-phased RV signals of the innermost, Neptune-like
companion~d, the new companion~e, the most massive planet~b, and the outermost
planet~c,  respectively. The open circles are for the RV measurements from
\cite{Butler2006}.  The error bars include the internal errors added in
quadrature to stellar jitter of 3.5~m/s. The next panel is for the Lomb-Scargle
periodogram of the RV data in the range of the short periods.   The last two
panels are for the periodogram of the residual signal after subtracting the
contribution of Jovian planets, and the histogram of $\Chi$ derived in the test
of scrambled residuals.
}

\figcaption[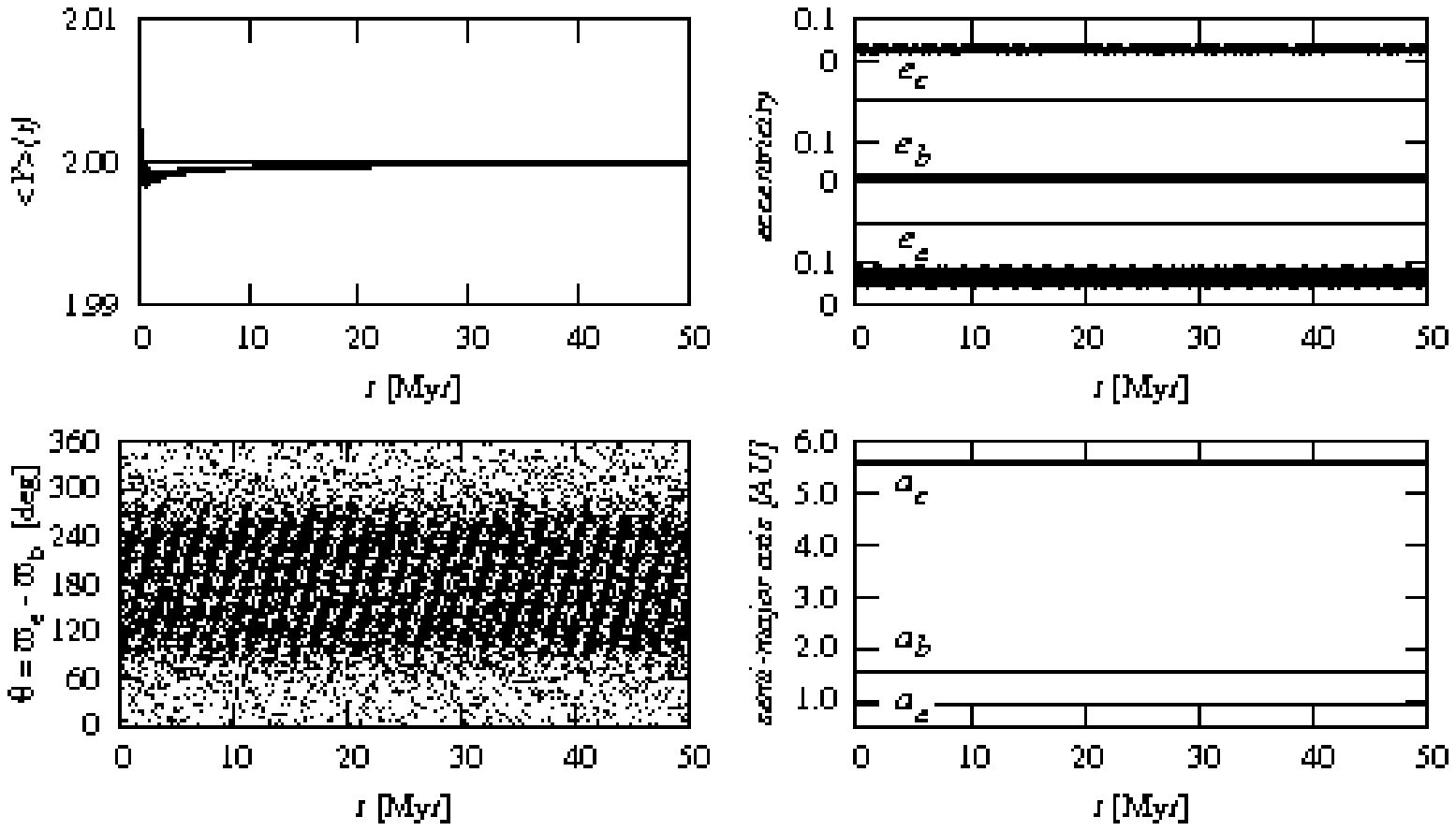]{\normalsize
Evolution of MEGNO (the top-left panel labeled by $\Ym$), and orbital elements
of the three-planet  configuration described by Fit~II given in Table~2  (only
Jovian planets are considered). A perfect convergence of MEGNO up to $\sim
50$~Myr indicates a rigorously stable solution over secular time scale. The
subsequent panels are for the eccentricities, the angle measuring the apsidal
anti-alignment of orbits e and~b,  and the semi-major axes. 
}

\figcaption[f8a.eps,f8b.eps]{\normalsize
The dynamical maps in the $(a_{\idm{b}},e_{\idm{b}})$-plane of the \muarae{}
system for the four-planet best Fit~II (Table~2).  The thin line marks the
collision curve for planets b and~e. See the caption to Fig.~\ref{fig:fig3} for 
an additional  explanation of the plots.  
}

\figcaption[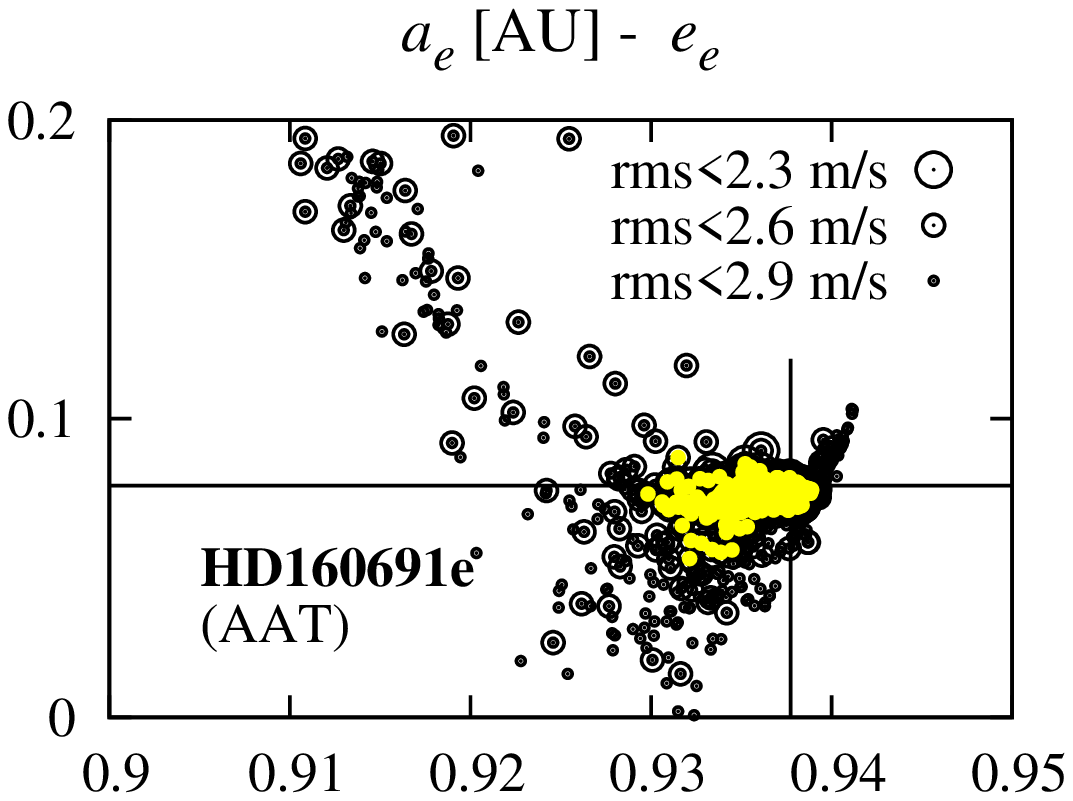,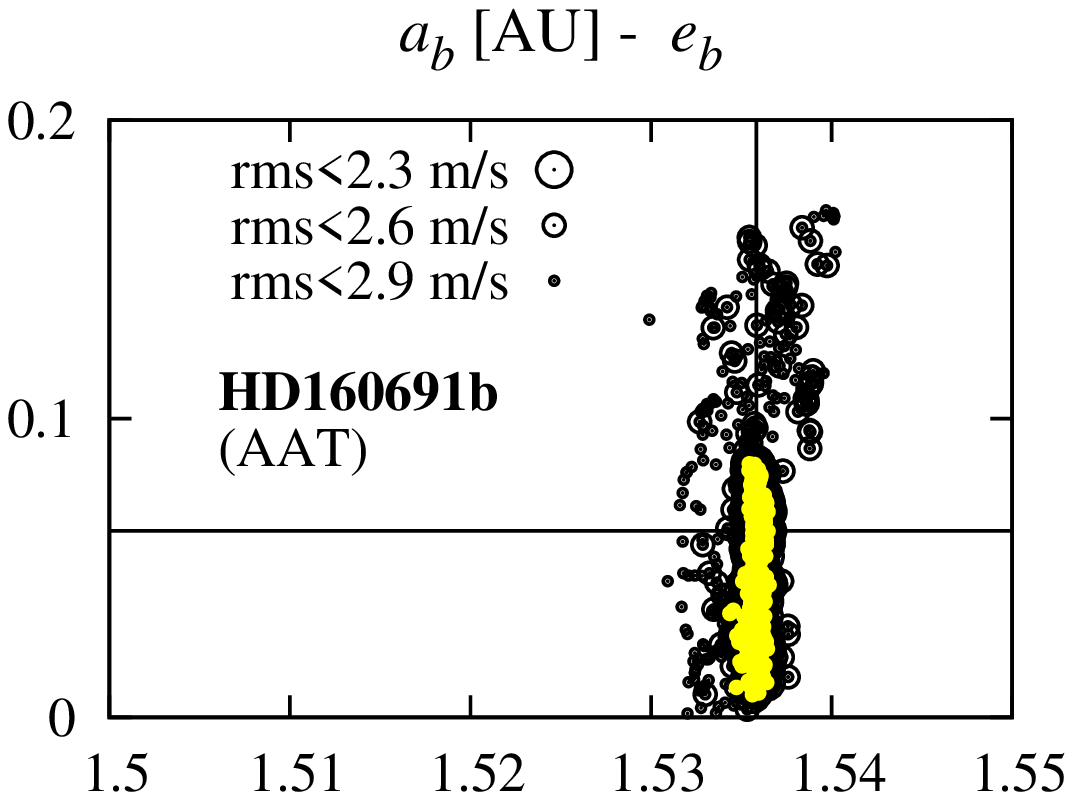,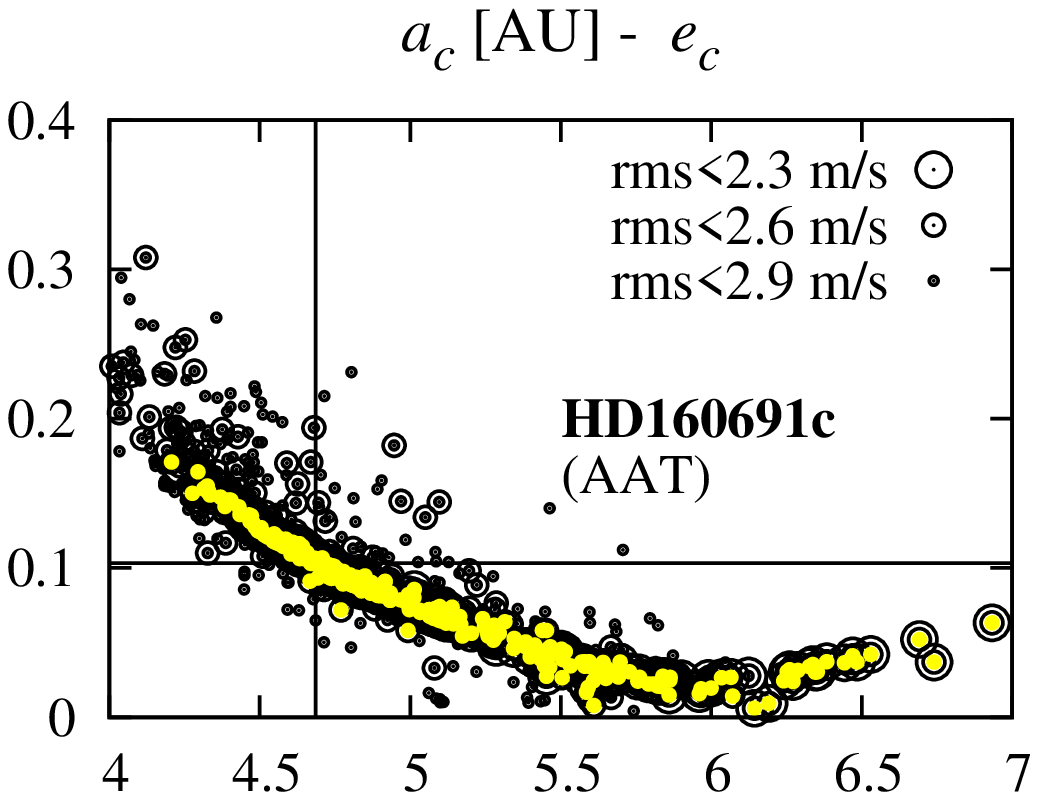,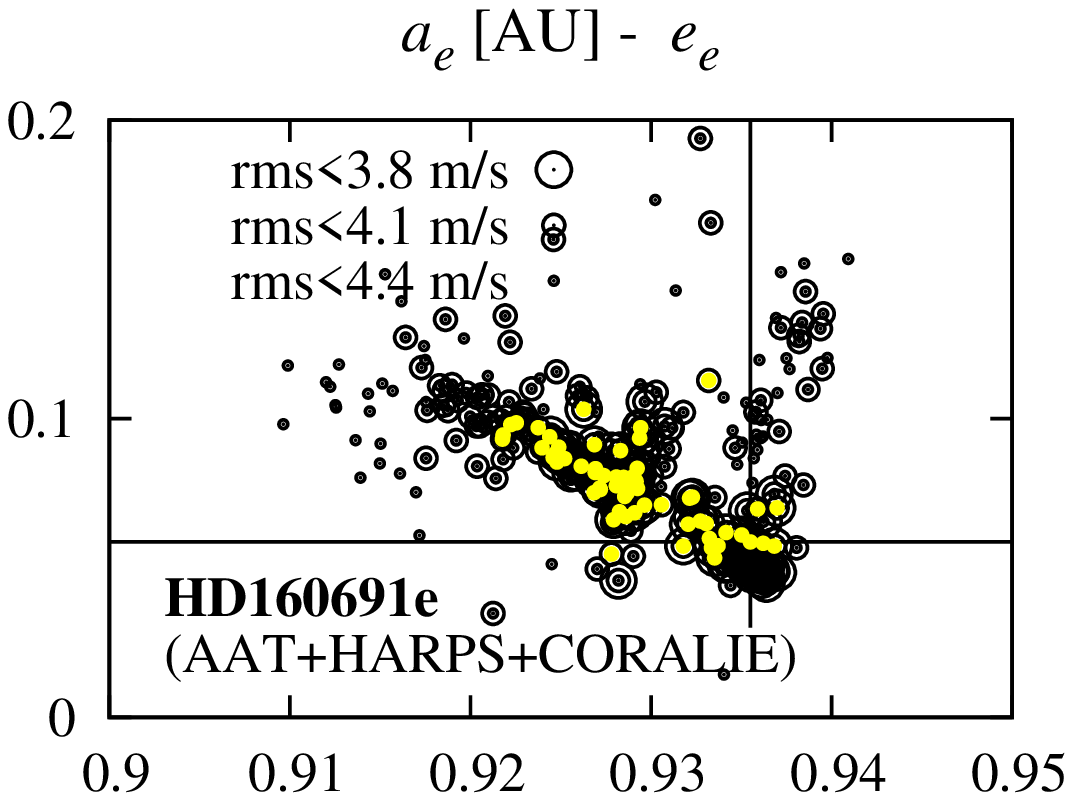,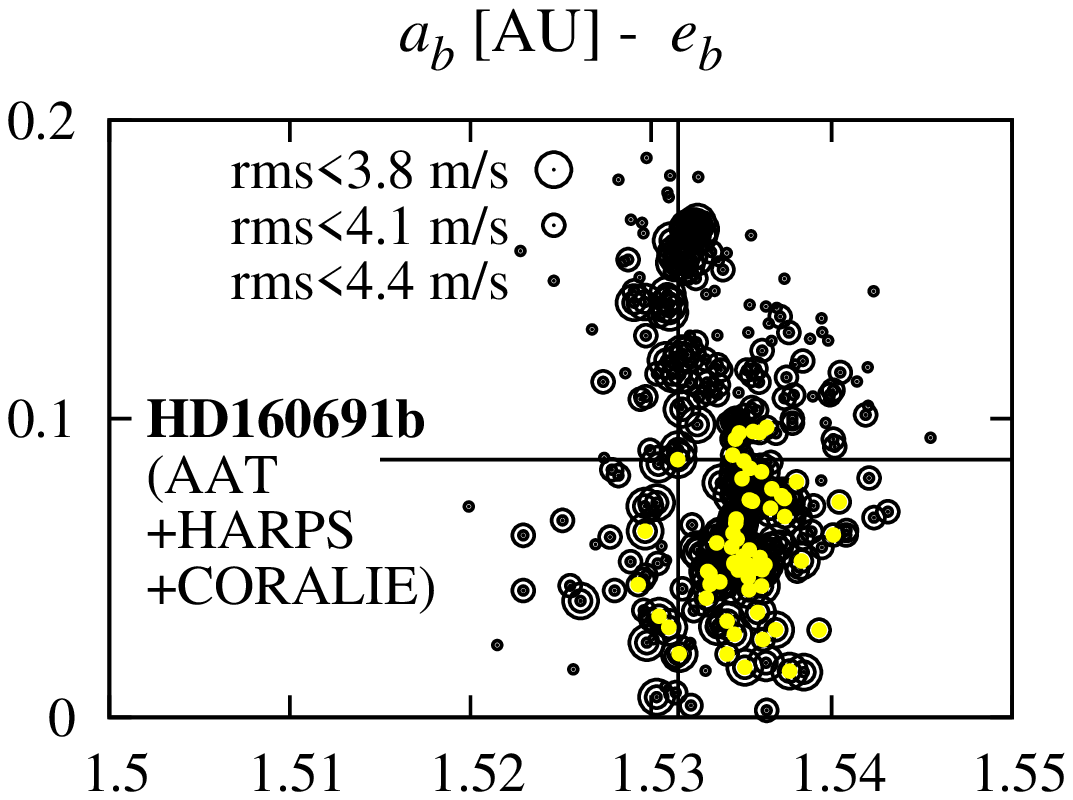,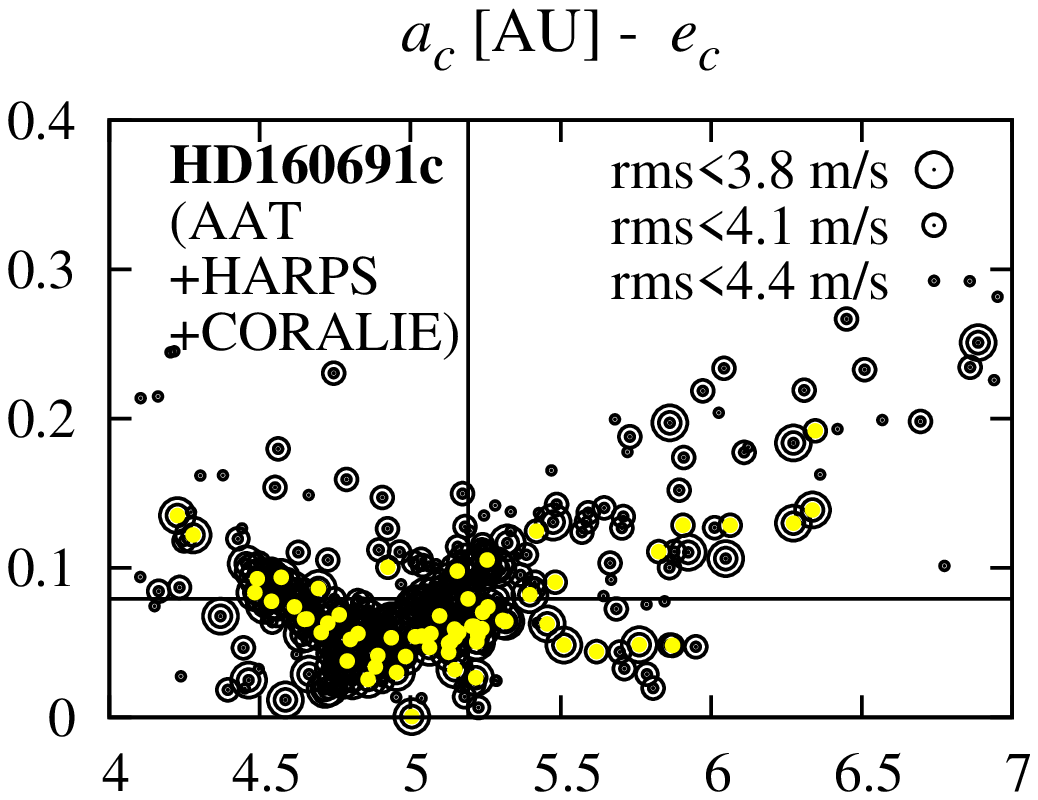,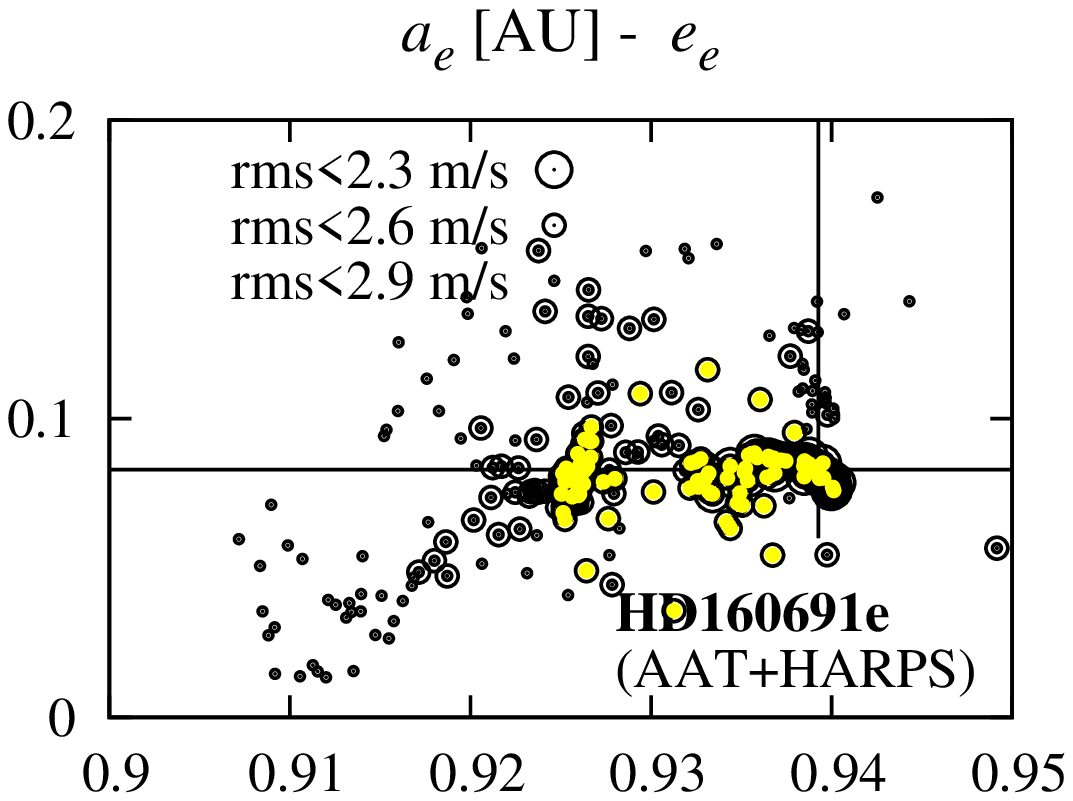,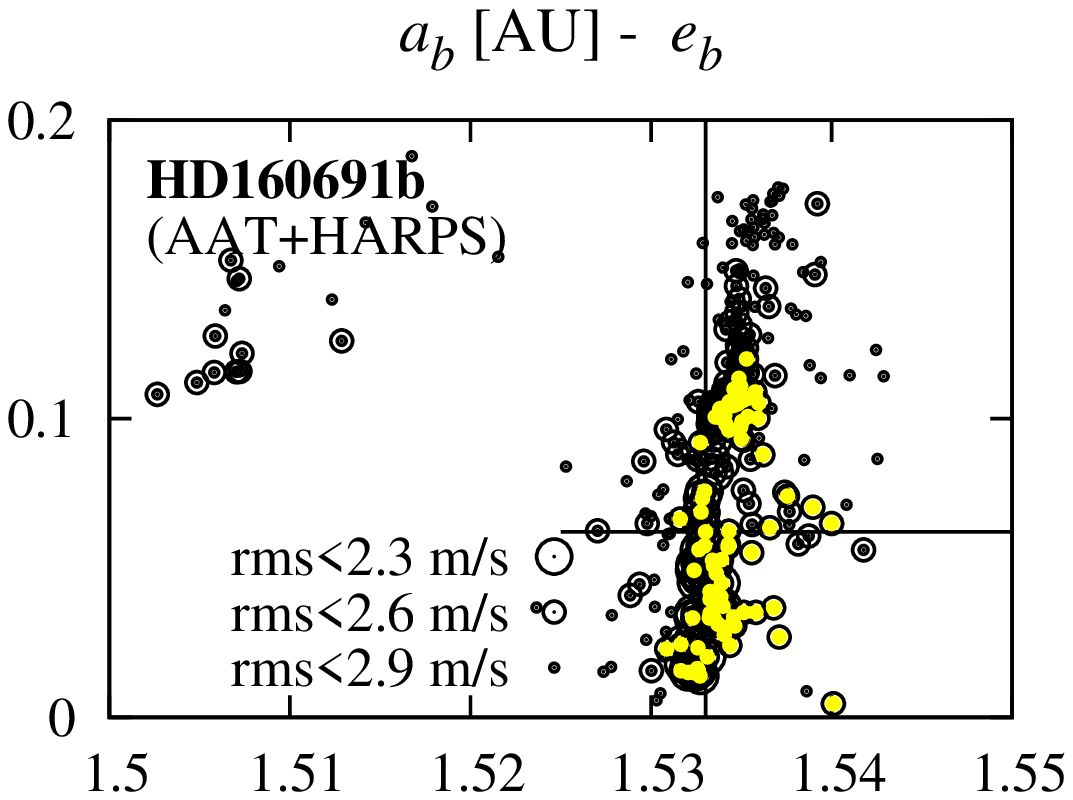,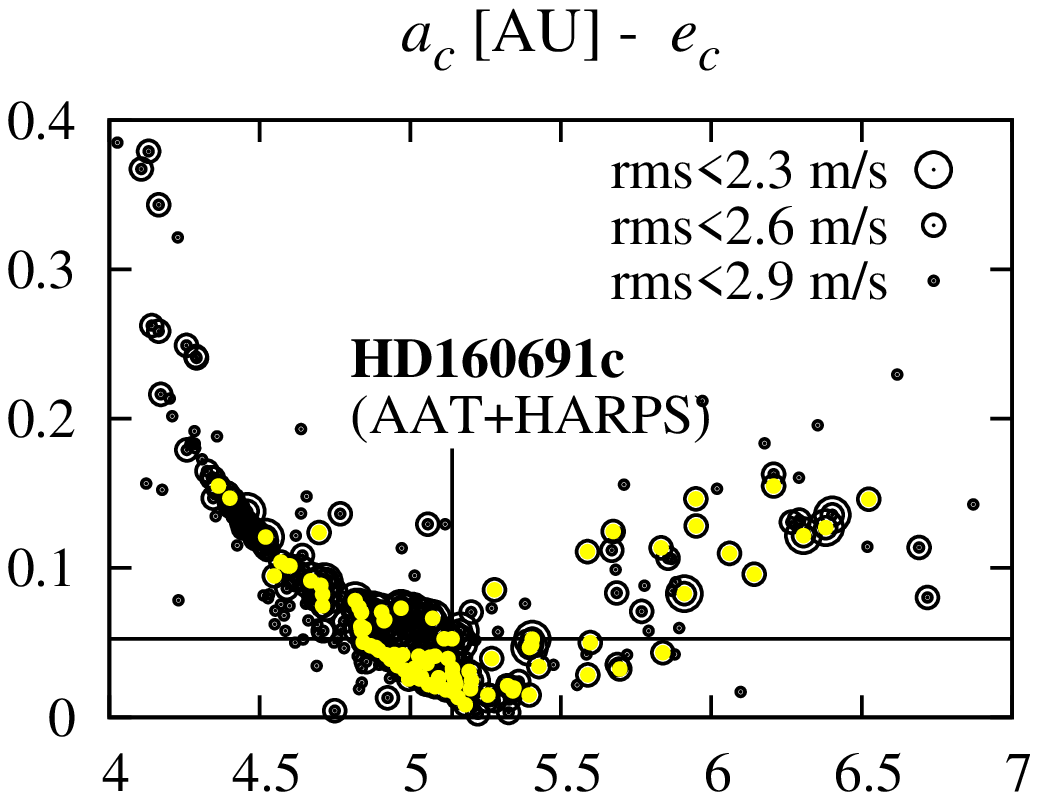]{
\normalsize
The statistics of the best, self-consistent Newtonian fits and their dynamical
stability. The subsequent panels are for the projections of the best fit,
osculating elements of the Jovian planets at the epoch of the relevant first
observation. The quality of fits, expressed by their rms, is marked by the size
of symbols (circles) and labeled in the plots. White (yellow--in the color
version of the figure) circles are for dynamically stable best-fit solutions.
The orbital stability is examined through MEGNO integrations over $\sim 3 \cdot
10^5 P_{\idm{c}}$, for all solutions with lowest rms (marked in the panels by
largest circles and labeled accordingly). Best fit-solutions marked as stable
have $|\Ym-2|<0.001$ at the end of the integration period. The upper row is for
the S1 data set  [the AAT measurements published by \citep{Butler2006}].  The
middle row is for the S2 data set (AAT+CORALIE+HARPS). The bottom row is for the
S3 data set (AAT+HARPS); note, that  in that case the rms was minimized instead
of $\Chi$.  Crossed lines marks the elements of the fits with lowest rms.
}

\figcaption[f10a.eps,f10b.eps,f10c.eps]{\normalsize
The dynamical maps in the $(a_{\idm{e}},e_{\idm{e}})$-plane computed for the
best fits to the S3 data set (including AAT and HARPS observations, see Fig.~9).
The maps are computed for the following  osculating elements of the Jovian
planets, given in terms of tuples
$(m~[M_{\idm{J}}], a~[\mbox{AU}], e, \omega~[\mbox{deg}], {\cal M}~[\mbox{deg}])$:
the left panel is for
 (0.430,    0.937,    0.086,  205.973,  282.203),
 (1.692,    1.532,    0.015,  314.000,   95.498),
 (1.704,    {\bf 4.702},    0.088,  161.445,   81.240);
the middle panel is for
 (0.549,    0.940,    0.077,  206.297,  298.451),
 (1.709,    1.534,    0.101,   47.902,    1.628),
 (1.808,    {\bf 4.969},    0.073,  128.199,  126.093);
and the right panel is for
 (0.485,    0.939,    0.085,  205.241,  289.819),
 (1.686,    1.533,    0.042,   41.633,    8.237),
 (2.643,    {\bf 6.306},    0.122,   23.664,  263.507),
respectively.
The thin line marks the collision line for planets b and~e. See the caption to
Fig.~\ref{fig:fig3} for an additional  explanation of the plots. 
}

\figcaption[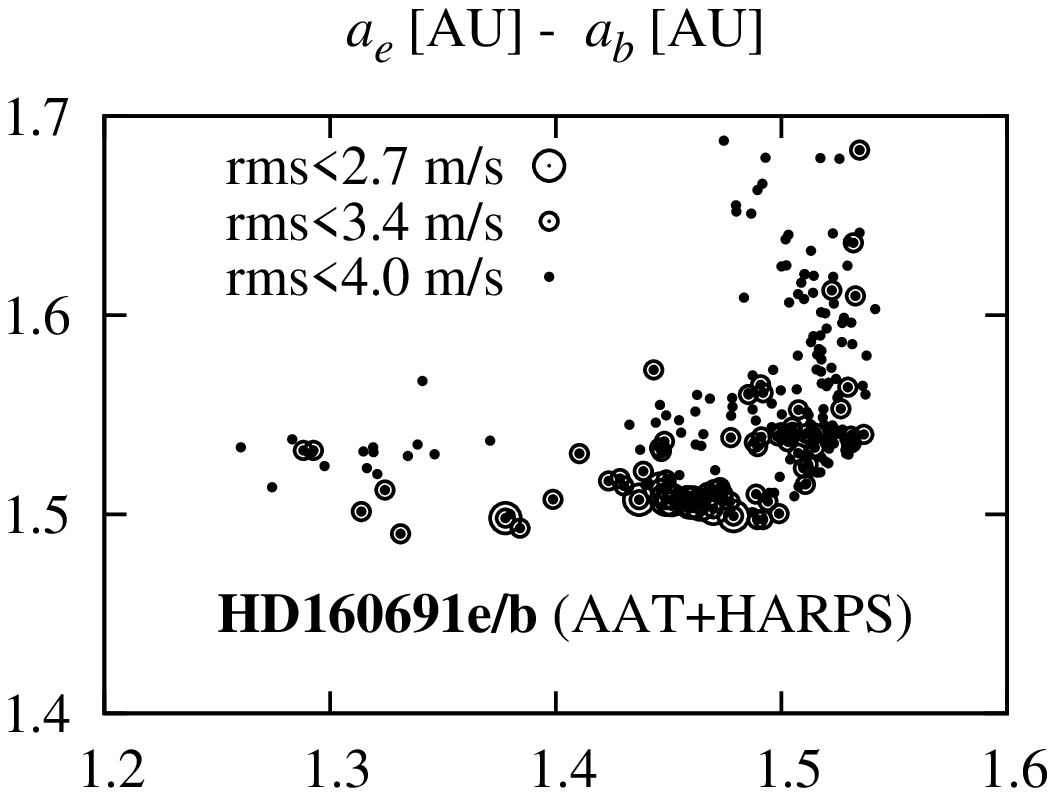,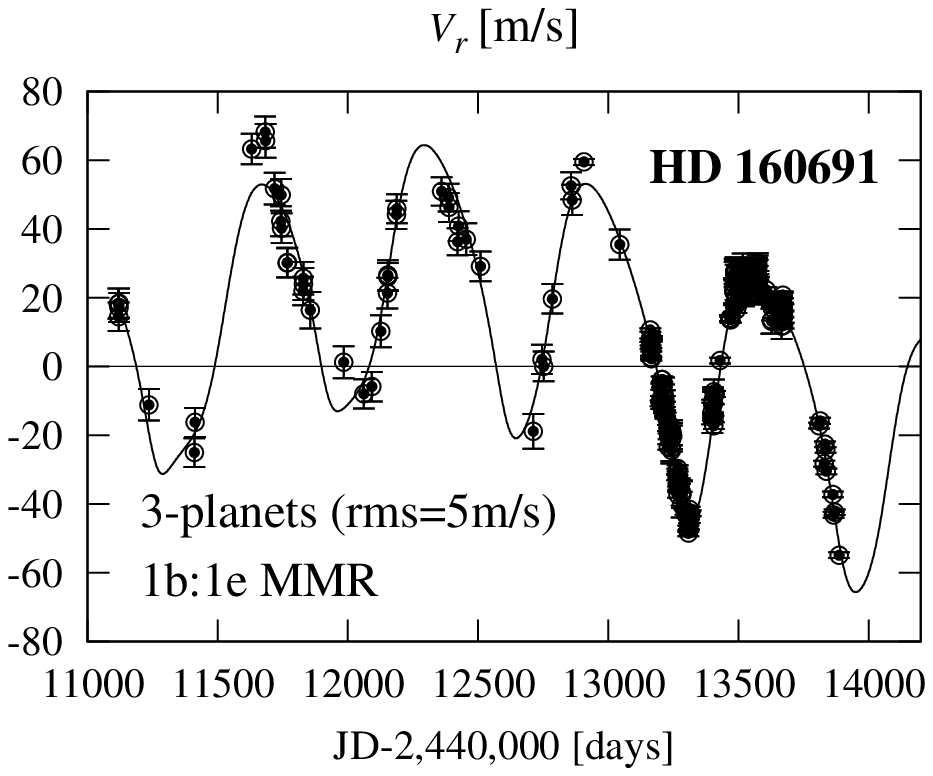,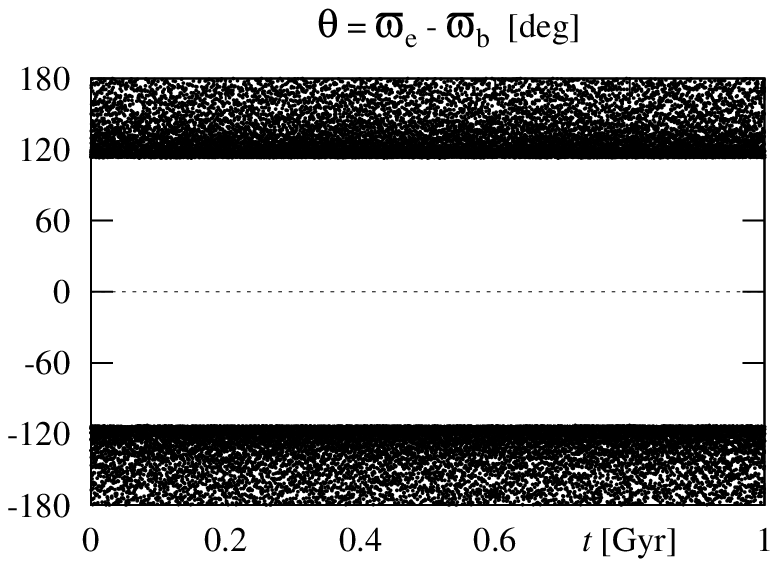,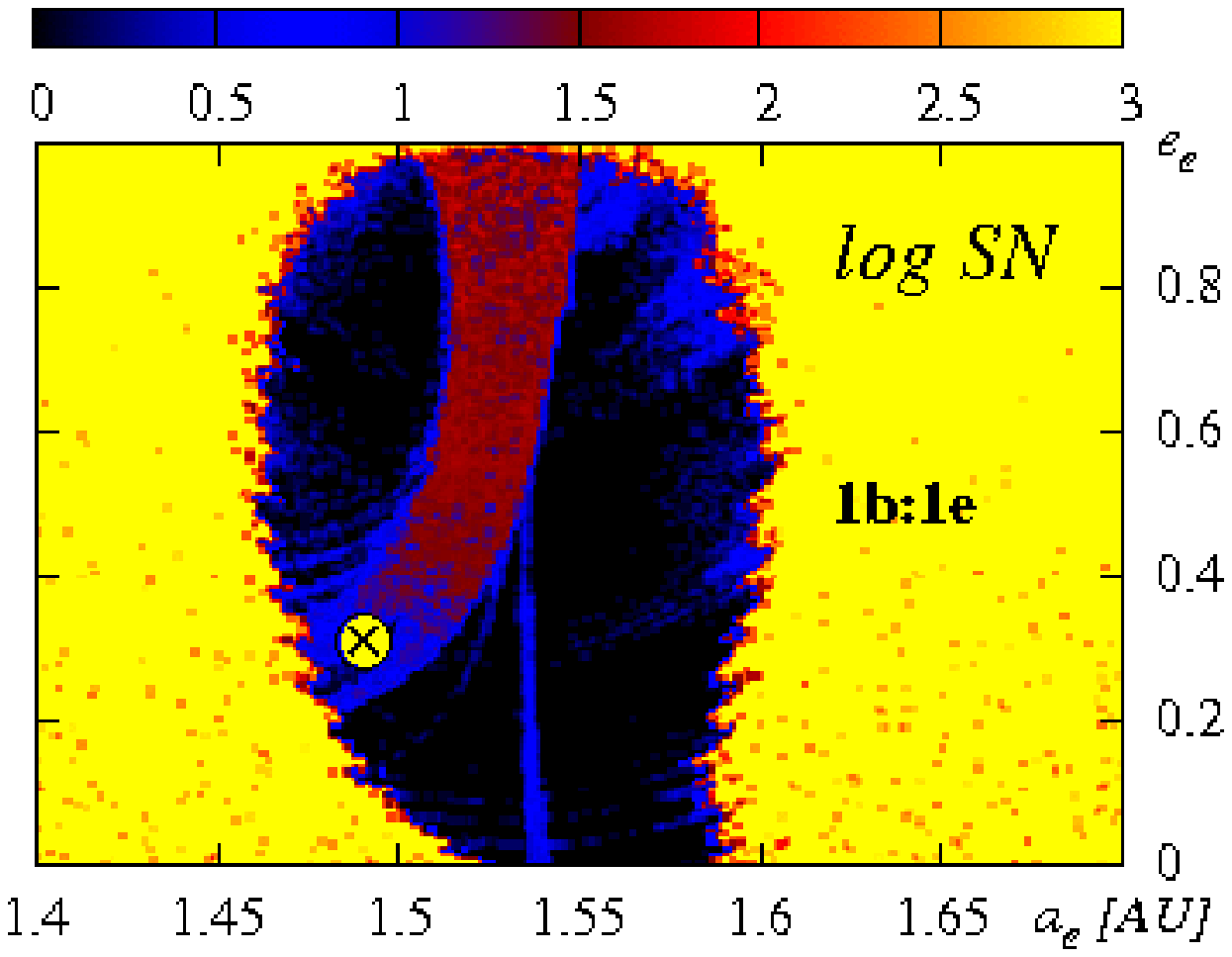]{\normalsize
The dynamical analysis of the orbital configuration of the \muarae{}  system
involving planets e-b in 1:1~MMR. The top-left panel is for the projections of
the best fits at the $(e_{\idm{b}},a_{\idm{e}}$)-plane of osculating elements,
as derived for the S3 data set (see the text for an explanation). The top-right
panel is for the synthetic curve of a stable solution, refined by GAMP
integrations over $\sim 10^5 P_{\idm{c}}$.  
The osculating elements of the 
three-Jovian system are:
(1.196,   1.52896,   0.3195, 180.208, 213.185),
(0.602,   1.49056,   0.305,  311.601,  69.827),
(1.939,   5.035,   0.016, 136.75, 110.0),
respectively, given in terms of tuples $(m~[M_{\idm{J}}],
a~[\mbox{AU}], e, \omega~[\mbox{deg}], {\cal M}~[\mbox{deg}])$, at the epoch of the
first observation, and the offsets are $V_1= -11.395$~m/s, $V_2=11.124$~m/s. The
bottom-left panel is for the time-evolution of
$\theta=\varpi_{\idm{e}}-\varpi_{\idm{b}}$, in the   best-fit configuration,
during 1~Gyr. The bottom-right panel is for the dynamical map in the vicinity of
the best fit; its position is marked by the crossed circle. 
}
%\clearpage

\setcounter{figure}{0}

%
% Figure 1
%
\begin{figure*}[!th]
   \centering
   \hbox{\includegraphics[]{f1.eps}}
   \caption{}
\label{fig:fig1}%
\end{figure*}
%
% Figure 2
%
\begin{figure*}
   \centering
   \hbox{\includegraphics[width=4in]{f2.eps}}
   \caption{}
\label{fig:fig2}%
\end{figure*}
%
% Figure 3
%
\begin{figure*}
   \centering
   \hbox{
   \hbox{\includegraphics[]{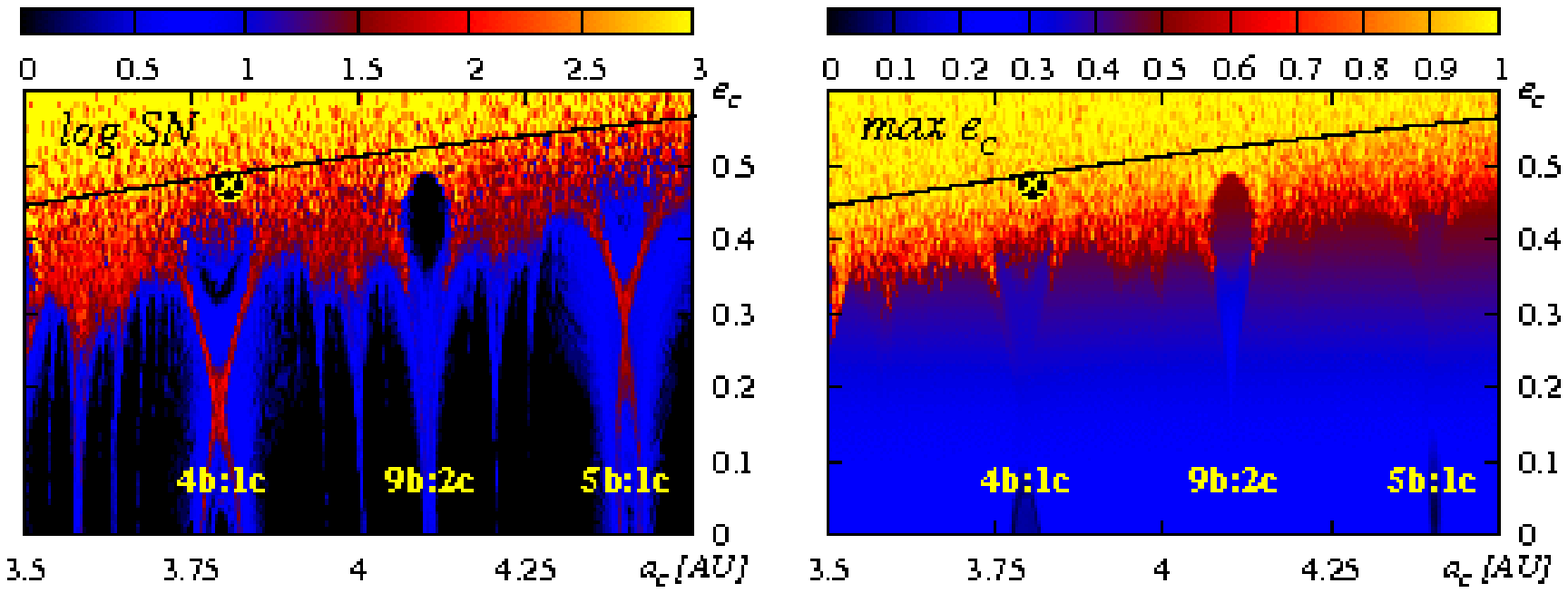}}
   }
   \caption{}
\label{fig:fig3}%
\end{figure*}
%
% Figure 4
%
\begin{figure*}
   \centering
   \hbox{
   \hbox{\includegraphics[]{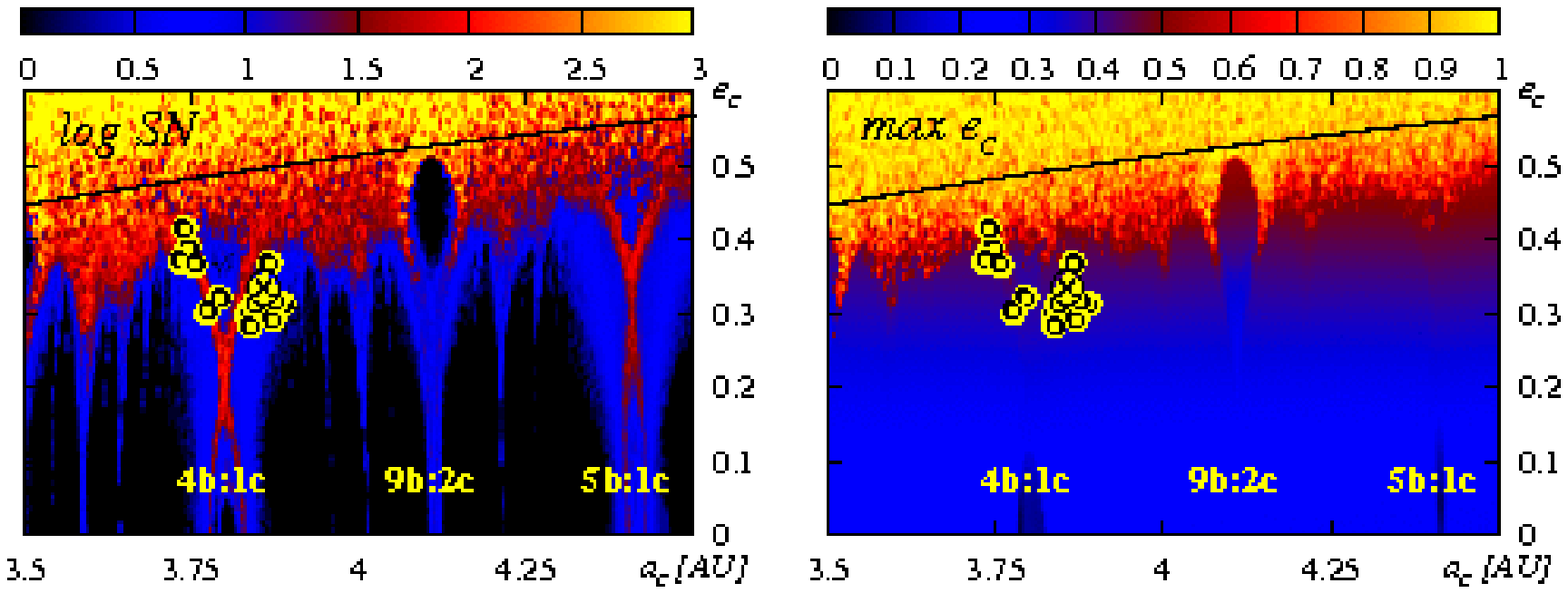}}
   }
   \caption{}
\label{fig:fig4}
\end{figure*}
%
% Figure 5
%
\begin{figure*}
   \centering
   \hbox{\includegraphics[]{f5.eps}}
   \caption{}
\label{fig:fig5}%
\end{figure*}
%
% Figure 6
%
\begin{figure*}
   \centering
   \hbox{
   \hbox{\includegraphics[width=3.3in]{f6a.eps}}
   \hbox{\includegraphics[width=3.3in]{f6b.eps}}
   }
   \hbox{
   \hbox{\includegraphics[width=3.3in]{f6c.eps}}
   \hbox{\includegraphics[width=3.3in]{f6d.eps}}
   }
   \hbox{
   \hbox{\includegraphics[width=3.3in]{f6e.eps}}
   \hbox{\includegraphics[width=3.3in]{f6f.eps}}
   }
   \hbox{
   \hbox{\includegraphics[width=3.3in]{f6g.eps}}
   \hbox{\includegraphics[width=3.3in]{f6h.eps}}
   }
   \caption{}
\label{fig:fig6}%
\end{figure*}
%
% Figure 7
%
\begin{figure*}
   \centering
   \hbox{\includegraphics[]{f7.eps}}
   \caption{}
\label{fig:fig7}%
\end{figure*}
%
% Figure 8 
%
\begin{figure*}
   \centering
   \hbox{
   \hbox{\includegraphics[]{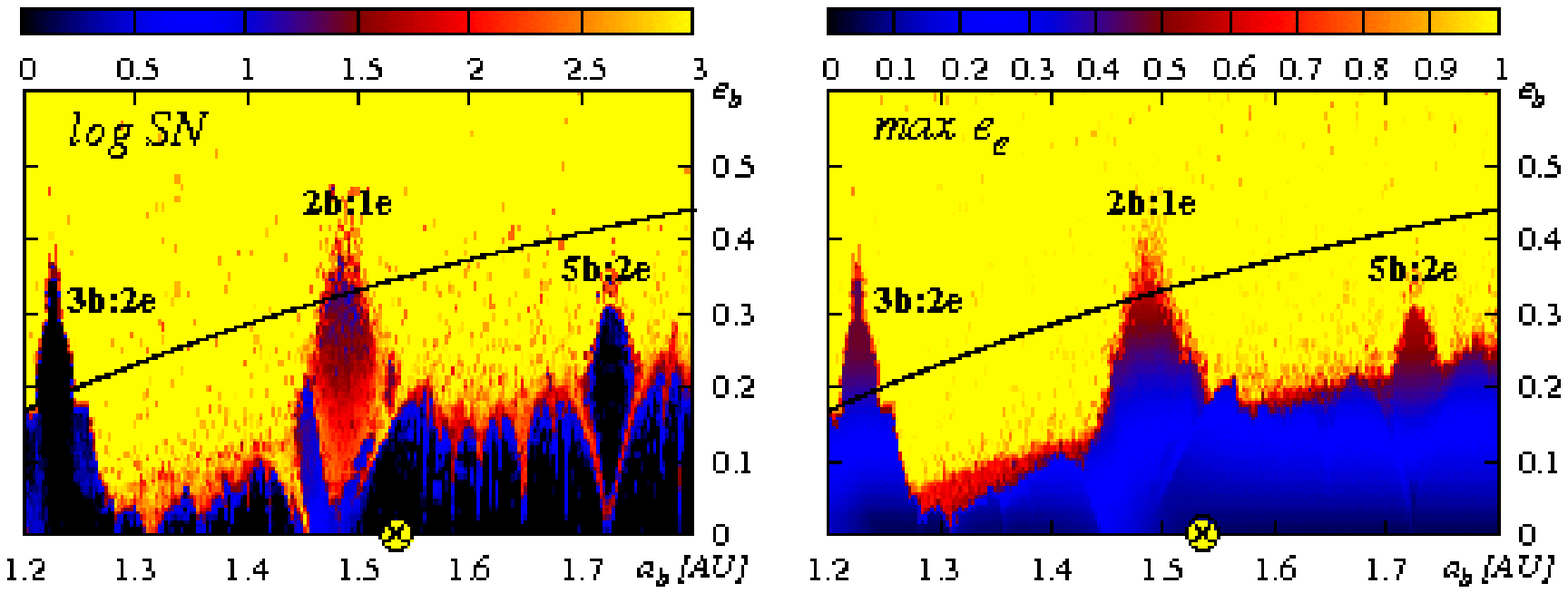}}
   }
   \caption{}
\label{fig:fig8}%
\end{figure*}
%
% Figure 9
%
\begin{figure*}
%   \centering  
   \vbox{
   \hbox{
   \kern-5em
   \hbox{\includegraphics[width=2.7in]{f9a.eps}}
   \kern-2em
   \hbox{\includegraphics[width=2.7in]{f9b.eps}}
   \kern-2em
   \hbox{\includegraphics[width=2.7in]{f9c.eps}}
   }
   \hbox{
   \kern-5em
   \hbox{\includegraphics[width=2.7in]{f9d.eps}}
   \kern-2em
   \hbox{\includegraphics[width=2.7in]{f9e.eps}}
   \kern-2em
   \hbox{\includegraphics[width=2.7in]{f9f.eps}}
   }
   \hbox{
   \kern-5em
   \hbox{\includegraphics[width=2.7in]{f9g.eps}}
   \kern-2em
   \hbox{\includegraphics[width=2.7in]{f9h.eps}}
   \kern-2em
   \hbox{\includegraphics[width=2.7in]{f9i.eps}}
   }
   }
   \caption{}                                    
\label{fig:fig9}%
\end{figure*}
%
% Figure 10
%
\begin{figure*}
   \centering
   \hbox{
   \kern-4em
   \hbox{\includegraphics[]{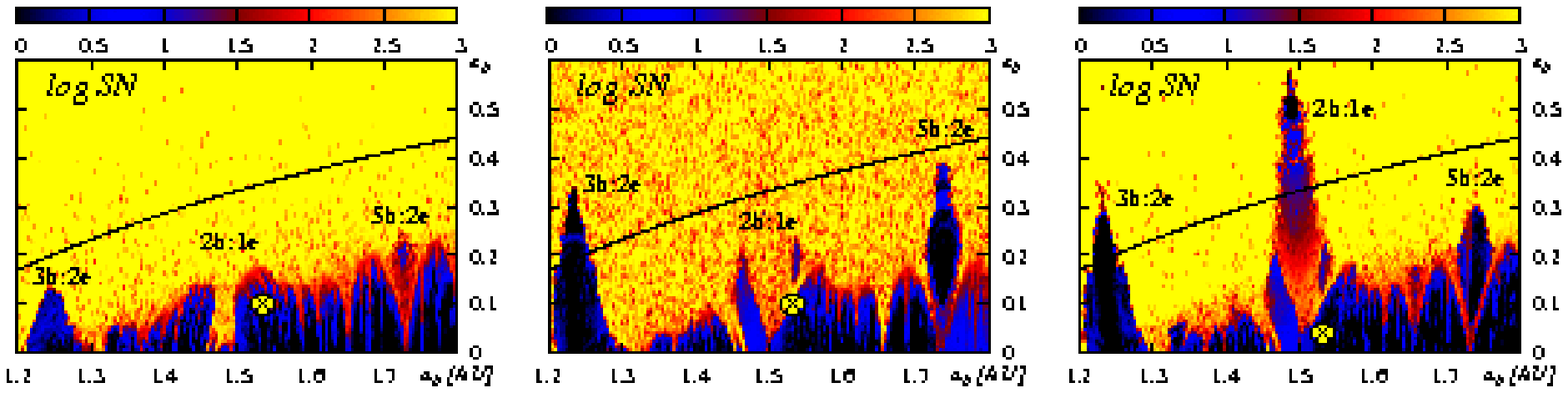}}
   }
   \caption{}
\label{fig:fig10}%
\end{figure*}
%
% Figure 11
%
\begin{figure*}
   \centering
   \hbox{
   \hbox{\includegraphics[width=3.25in]{f11a.eps}}
   \hbox{\includegraphics[width=3.05in]{f11b.eps}}
   }
   \hbox{
   \hbox{\includegraphics[width=3.05in]{f11c.eps}}
   \kern2em\hbox{\includegraphics[width=3.05in]{f11d.eps}}
   }
   \caption{}
\label{fig:fig11}%
\end{figure*}

%
%------------------------------------------------------------------------------
%

\begin{thebibliography}{31}
%
%------------------------------------------------------------------------------
%
\expandafter\ifx\csname natexlab\endcsname\relax\def\natexlab#1{#1}\fi

\bibitem[{{Arnold}(1978)}]{Arnold1978}
{Arnold}, V.~I. 1978, {Mathematical methods of classical mechanics} (Graduate
  texts in mathematics, New York: Springer, 1978)


\bibitem[{{Beaug{\'e}} {et~al.}(2003){Beaug{\'e}}, {Ferraz-Mello}, \&
  {Michtchenko}}]{Beauge2003}
{Beaug{\'e}}, C., {Ferraz-Mello}, S., \& {Michtchenko}, T.~A. 2003, \apj, 593,
  1124

\bibitem[{{Bevington} \& {Robinson}(2003)}]{Bevington2003}
{Bevington}, P.~R. \& {Robinson}, D.~K. 2003, {Data reduction and error
  analysis for the physical sciences} (McGraw-Hill)

\bibitem[{{Bouchy} {et~al.}(2005){Bouchy}, {Bazot}, {Santos}, {Vauclair}, \&
  {Sosnowska}}]{Bouchy2005}
{Bouchy}, F., {Bazot}, M., {Santos}, N.~C., {Vauclair}, S., \& {Sosnowska}, D.
  2005, \aap, 440, 609

\bibitem[{{Butler} {et~al.}(2001){Butler}, {Tinney}, {Marcy}, {Jones}, {Penny},
  \& {Apps}}]{Butler2001}
{Butler}, R.~P., {Tinney}, C.~G., {Marcy}, G.~W., {Jones}, H.~R.~A., {Penny},
  A.~J., \& {Apps}, K. 2001, \apj, 555, 410

\bibitem[{{Butler} {et~al.}(2004){Butler}, {Vogt}, {Marcy}, {Fischer},
  {Wright}, {Henry}, {Laughlin}, \& {Lissauer}}]{Butler2004}
{Butler}, R.~P., {Vogt}, S.~S., {Marcy}, G.~W., {Fischer}, D.~A., {Wright},
  J.~T., {Henry}, G.~W., {Laughlin}, G., \& {Lissauer}, J.~J. 2004, \apj, 617,
  580

\bibitem[{{Butler} {et~al.}(2006){Butler}, {Wright}, {Marcy}, {Fischer},
  {Vogt}, {Tinney}, {Jones}, {Carter}, {Johnson}, {McCarthy}, \&
  {Penny}}]{Butler2006}
{Butler}, R.~P., {Wright}, J.~T., {Marcy}, G.~W., {Fischer}, D.~A., {Vogt},
  S.~S., {Tinney}, C.~G., {Jones}, H.~R.~A., {Carter}, B.~D., {Johnson}, J.~A.,
  {McCarthy}, C., \& {Penny}, A.~J. 2006, \apj, 646, 505
  
  
\bibitem[{{Cincotta} \& {Sim{\' o}}(2000)}]{Cincotta2000}
{Cincotta}, P.~M. \& {Sim{\' o}}, C. 2000, \aaps, 147, 205

\bibitem[{{Charbonneau}(1995)}]{Charbonneau1995}
{Charbonneau}, P. 1995, \apjs, 101, 309

\bibitem[{{Ford} \& {Gaudi}(2006)}]{Ford2006}
{Ford}, E. \& {Gaudi}, B.~S.  2006, astroph/0609298


\bibitem[{{Go{\' z}dziewski} \& {Konacki}(2004)}]{Gozdziewski2004}
{Go{\' z}dziewski}, K. \& {Konacki}, M. 2004, \apj, 610, 1093

\bibitem[{{Go{\' z}dziewski} \& {Konacki}(2006)}]{Gozdziewski2006c}
{Go{\' z}dziewski}, K. \& {Konacki}, M. 2006, \apj, 647, 573

\bibitem[{{Go{\' z}dziewski} {et~al.}(2003){Go{\' z}dziewski}, {Konacki}, \&
  {Maciejewski}}]{Gozdziewski2003e}
{Go{\' z}dziewski}, K., {Konacki}, M., \& {Maciejewski}, A.~J. 2003, \apj, 594

\bibitem[{{Go{\' z}dziewski} {et~al.}(2005){Go{\' z}dziewski}, {Konacki}, \&
  {Maciejewski}}]{Gozdziewski2005a}
---. 2005, \apj, 622, 1136

\bibitem[{{Go{\'z}dziewski} {et~al.}(2006){Go{\'z}dziewski}, {Konacki}, \&
  {Maciejewski}}]{Gozdziewski2006a}
{Go{\'z}dziewski}, K., {Konacki}, M., \& {Maciejewski}, A.~J. 2006, \apj, 645,
  688

\bibitem[{{Go{\'z}dziewski} {et~al.}(2005){Go{\'z}dziewski}, {Konacki}, \&
  {Wolszczan}}]{Gozdziewski2005b}
{Go{\'z}dziewski}, K., {Konacki}, M., \& {Wolszczan}, A. 2005, \apj, 619, 1084

\bibitem[{{Go{\'z}dziewski} \& {Migaszewski}(2006)}]{Gozdziewski2006b}
{Go{\'z}dziewski}, K. \& {Migaszewski}, C. 2006, \aap, 449, 1219

\bibitem[{{Jones} {et~al.}(2002){Jones}, {Butler}, {Marcy}, {Tinney}, {Penny},
  C., \& B.}]{Jones2002a}
{Jones}, H., {Butler}, P., {Marcy}, G., {Tinney}, C., {Penny}, A., C., M., \&
  B., C. 2002, MNRAS, 337, 1170

\bibitem[{{Ji} {et~al.}(2003)}]{Ji2003}
Ji, J. {et~al.} 2003, Celestial Mech. Dyn. Astr, 74, 113

\bibitem[{{Laskar}(2000)}]{Laskar2000}
{Laskar}, J. 2000, Physical Review Letters, 84, 3240

\bibitem[{{Laughlin} \& {Chambers}(2001)}]{Laughlin2001}
{Laughlin}, G. \& {Chambers}, J.~E. 2001, ApJ, 551, L109

\bibitem[{{Lee} {et~al.}(2006){Lee}, {Butler}, {Fischer}, {Marcy}, \&
  {Vogt}}]{Lee2006}
{Lee}, M.~H., {Butler}, R.~P., {Fischer}, D.~A., {Marcy}, G.~W., \& {Vogt},
  S.~S. 2006, \apj, 641, 1178

\bibitem[{{Lee} \& {Peale}(2003)}]{Lee2003}
{Lee}, M.~H. \& {Peale}, S.~J. 2003, \apj, 592, 1201

\bibitem[{{Lovis} {et~al.}(2006){Lovis}, {Mayor}, {Pepe}, {Alibert}, {Benz},
  {Bouchy}, {Correia}, {Laskar}, {Mordasini}, {Queloz}, {Santos}, {Udry},
  {Bertaux}, \& {Sivan}}]{Lovis2006}
{Lovis}, C., {Mayor}, M., {Pepe}, F., {Alibert}, Y., {Benz}, W., {Bouchy}, F.,
  {Correia}, A.~C.~M., {Laskar}, J., {Mordasini}, C., {Queloz}, D., {Santos},
  N.~C., {Udry}, S., {Bertaux}, J.-L., \& {Sivan}, J.-P. 2006, \nat, 441, 305

\bibitem[{{Mayor} {et~al.}(2004){Mayor}, {Udry}, {Naef}, {Pepe}, {Queloz},
  {Santos}, \& {Burnet}}]{Mayor2004}
{Mayor}, M., {Udry}, S., {Naef}, D., {Pepe}, F., {Queloz}, D., {Santos}, N.~C.,
  \& {Burnet}, M. 2004, \aap, 415, 391

\bibitem[{{McArthur} {et~al.}(2004){McArthur}, {Endl}, {Cochran}, {Benedict},
  {Fischer}, {Marcy}, {Butler}, {Naef}, {Mayor}, {Queloz}, {Udry}, \&
  {Harrison}}]{McArthur2005}
{McArthur}, B.~E., {Endl}, M., {Cochran}, W.~D., {Benedict}, G.~F., {Fischer},
  D.~A., {Marcy}, G.~W., {Butler}, R.~P., {Naef}, D., {Mayor}, M., {Queloz},
  D., {Udry}, S., \& {Harrison}, T.~E. 2004, \apjl, 614, L81

\bibitem[{McCarthy {et~al.}(2004)}]{McCarthy2004}
McCarthy, C. {et~al.} 2004, \apj, 617, 575

\bibitem[{{Michtchenko} \& {Ferraz-Mello}(2001)}]{Michtchenko2001}
{Michtchenko}, T. \& {Ferraz-Mello}, S. 2001, ApJ, 122, 474

\bibitem[{Nauenberg {M.}(2002)}]{Nauenberg2002}
Nauenberg, C. 2002, \apj, 124, 233

\bibitem[{{Pepe} {et~al.}(2006)}]{Pepe2006}
Pepe, F. {et~al} 2006, astro-ph/0608396

\bibitem[{{Press} {et~al.}(1992){Press}, {Teukolsky}, Vetterling, \&
  {Flannery}}]{Press1992}
{Press}, W.~H., {Teukolsky}, S.~A., Vetterling, W.~T., \& {Flannery}, B.~P.
  1992, Numerical Recipes in C. The Art of Scientific Computing (Cambridge
  Univ. Press)

\bibitem[{{Psychoyos} \& {Hadjidemetriou}(2005)}]{Psychoyos2005}
{Psychoyos}, D. \& {Hadjidemetriou}, J.~D. 2005, Celestial Mech. Dyn. Astr.,
92, 135

\bibitem[{{Rivera} \& {Lissauer}(2001)}]{Rivera2001}
{Rivera}, E.~J. \& {Lissauer}, J.~J. 2001, ApJ, 402, 558

\bibitem[{{Rivera} {et~al.}(2005){Rivera}, {Lissauer}, {Butler}, {Marcy},
  {Vogt}, {Fischer}, {Brown}, {Laughlin}, \& {Henry}}]{Rivera2005}
{Rivera}, E.~J., {Lissauer}, J.~J., {Butler}, R.~P., {Marcy}, G.~W., {Vogt},
  S.~S., {Fischer}, D.~A., {Brown}, T.~M., {Laughlin}, G., \& {Henry}, G.~W.
  2005, \apj, 634, 625

\bibitem[{Santos {et~al.}(2004)}]{Santos2004}
Santos, N.~C. {et~al.} 2004, \aap, 426, L19

\bibitem[{{Smart}(1949)}]{Smart1949}
{Smart}, W.~M. 1949, {Text-Book on Spherical Astronomy} (Cambridge Univ. Press)

\bibitem[{{Tinney} {et~al.}(2006){Tinney}, {Butler}, {Marcy}, {Jones}~{Penny},
  {McCarthy}, {Carter}, \& {Fischer}}]{Tinney2006}
{Tinney}, C.~G., {Butler}, R.~P., {Marcy}, G.~W., {Jones}~{Penny}, A.~J.,
  {McCarthy}, C., {Carter}, B.~D., \& {Fischer}, D. 2006, \apj, 647, 594

\bibitem[{{Vogt} {et~al.}(2005){Vogt}, {Butler}, {Marcy}, {Fischer}, {Henry},
  {Laughlin}, {Wright}, \& {Johnson}}]{Vogt2005}
{Vogt}, S.~S., {Butler}, R.~P., {Marcy}, G.~W., {Fischer}, D.~A., {Henry},
  G.~W., {Laughlin}, G., {Wright}, J.~T., \& {Johnson}, J.~A. 2005, \apj, 632,
  638

\bibitem[{{Wright}(2005)}]{Wright2005}
{Wright}, J.~T. 2005, \pasp, 117, 657

\end{thebibliography}
\end{document}